\documentclass[fleqn,usenatbib]{mnras}

\usepackage{newtxtext,newtxmath}

\usepackage[T1]{fontenc}
\usepackage{ae,aecompl}

\usepackage{graphicx}	
\usepackage{amsmath}	
\usepackage{amssymb}	
\usepackage{natbib}
\title[\textsc{LineStacker}: A spectral line stacking tool]{\textsc{LineStacker}: A spectral line stacking tool for interferometric data}

\author[J.-B. Jolly et al.]{
Jean-Baptiste Jolly,$^{1}$\thanks{E-mail: jean.jolly@chalmers.se}
Kirsten K. Knudsen,$^{1}$
and Flora Stanley$^{1}$
\\
$^{1}$Department of Space, Earth and Environment, Chalmers University of Technology, Onsala Space Observatory, SE-439 92 Onsala, Sweden\\
}

\date{Accepted XXX. Received YYY; in original form ZZZ}

\pubyear{2020}

\begin{document}
\label{firstpage}
\pagerange{\pageref{firstpage}--\pageref{lastpage}}
\maketitle

\begin{abstract}
\textsc{LineStacker} is a new open access and open source tool for stacking of spectral lines in interferometric data. \textsc{LineStacker} is an ensemble of CASA tasks, and can stack both 3D cubes or already extracted spectra. The algorithm is tested on increasingly complex simulated data sets, mimicking Atacama Large Millimeter/submillimeter Array and Karl G. Jansky Very Large Array observations of [C\,\textsc{ii}] and CO(3-2) emission lines, from $z\sim7$ and $z\sim4$ galaxies respectively.
We find that the algorithm is very robust, successfully retrieving the input parameters of the stacked lines in all cases with an accuracy $\gtrsim90$\%. However, we distinguish some specific situations showcasing the intrinsic limitations of the method. Mainly that high uncertainties on the redshifts ($\Delta z > 0.01$) can lead to poor signal to noise ratio improvement, due to lines being stacked on shifted central frequencies. Additionally we give an extensive description of the embedded statistical tools included in \textsc{LineStacker}: mainly bootstrapping, rebinning and subsampling. Velocity rebinning {is applied on the data before stacking and} proves necessary when studying line profiles, in order to avoid artificial spectral features in the stack. Subsampling is useful to sort the stacked sources, allowing to find a subsample maximizing the searched parameters, while bootstrapping allows to detect inhomogeneities in the stacked sample. \textsc{LineStacker} is a useful tool for extracting the most from spectral observations of various types. 
\end{abstract}





\section{Introduction} \label{sec:intro}
One of the great challenges in astronomy, and more especially in the field of galaxy evolution, comes from the tendency to look primarily at the brightest sources of a given galaxy population. The further we look the fewer intrinsically faint objects are observable, posing a real difficulty in studying faint, low-mass, galaxy properties, or faint tracer of physical and chemical processes. In order to draw an accurate description of a galaxy population it is necessary to study representative samples, including the faint/undetected sources. One method that can be used for this purpose is stacking.

Stacking was first developed for optical data \citep{CadyBates1980} to determine the average properties of otherwise undetected sources and has, since then, been frequently used for studies at many different wavelength ranges \citep[e.g.][]{Kirsten2005,Hickox2007,Hickox2009,Karim2011,Chen2013,Lindroos2015,Lindroos2016,Stanley2017}. While radio and mm wavelengths observations are essential to study the gas content of high-redshift galaxies, arcsecond and sub-arcsecond angular resolution can only be obtained when doing interferometric observations.  However, interferometry is not a direct imaging technique, but instead samples the Fourier transform of the brightness distribution of the sources being observed. Therefore, the produced images are a model representation of the actual data \citep[see e.g.][]{Thompson2001}. While these models are well understood they often lead to the generation of artifacts, making stacking analysis of interferometric data less straightforward \citep{Lindroos2015}

Stacking of radio and mm wavelength interferometric data has mainly been done for continuum data \citep[e.g.][]{Karim2011,Decarli2014,ikarashi15,Lindroos2016,Lindroos2018,Stanley2017}, it has occasionally been done for spectral lines as well  \citep[e.g.][]{Murray2014,Decarli2018QSO,Bischetti2018,Stanley2019,Fujimoto2019}. However, no general open access tool nor any thorough study of the spectral stacking method has yet been published.

In this paper we describe the functionalities and performances of a new tool for stacking spectral lines: \textsc{LineStacker}. \textsc{LineStacker} is an ensemble of CASA tasks and it allows stacking of spectral cubes in the image-plane. Its main contribution is the stacking algorithm, but also includes embedded statistical tools for further analysis of the stacked data and optimisation of the stacked signal. We also demonstrate the performances and capabilities of \textsc{LineStacker}, by testing it on increasingly complex simulated data-sets, that mimic mm- and radio observation of emission lines from high-redshift galaxies. The tests are performed on both high and low signal-to-noise ratio (SNR) cases. The high SNR data sets test the reliability of the algorithm in a near-ideal case, while the low SNR data sets are used to verify the efficiency of noise reduction. 

In \cite{Stanley2019} we used \textsc{LineStacker} to perform a spectral stacking analysis to search for faint outflow signatures in a sample of $z\sim6$ quasars. We used the main algorithm and accompanying tools presented in this paper, 
on a sample of 26 quasars with detected [C\,\textsc{ii}] emission.
Our work demonstrated the utility of \textsc{LineStacker} as a spectral stacking tool, when searching for faint emission at high redshift.

In sections \ref{stackAlgo} and \ref{sec:stat} we give a complete description of \textsc{LineStacker}, fully characterizing both the main algorithm and the embedded tools. 
In section \ref{simu} we describe each simulated data set in detail. 
In section \ref{sec:results} we give the results from our stacking analysis on the simulated data-sets. We discuss the results of the analysis, and review possible outlooks in section \ref{discussion}. Finally section \ref{ccl} outlines the conclusions of this study.


\section{\textsc{LineStacker}} \label{stackAlgo}

\textsc{LineStacker} is an assembly of CASA tasks allowing stacking of data cubes, specifically cubes with two spatial dimensions and a frequency/velocity dimension. It has been developed specifically to stack spectral lines, with the capability to take into account varying redshift/central-velocities across the sample. In addition, embedded analysis tools are included within \textsc{LineStacker}, for further analyis of the stack results and sample. \textsc{LineStacker} is an extension of \textsc{Stacker}, \citep{Lindroos2015} a tool built for stacking continuum interferometric data. While \textsc{Stacker} allowed direct visibility stacking, the visibility stacking extension of \textsc{LineStacker} is still under construction.

\subsection{Main algorithm} \label{mainAlgo}

When stacking cubes, every source is stacked pixel to pixel, spectral bin to spectral bin. Spatial stacking positions as well as observed central frequency -- or rest-frame frequency and redshift -- of the sources are needed prior to stacking. The position of the source can typically be obtained through continuum observations, or, more generally, through prior observations at other wavelength.
Subimages of $N\times N$ pixels are stacked, centered on the stacking position. Similarly only a subset of the total number of spectral bins, centered on the estimated central frequency of the line in the observer frame, are stacked (note that all the spectral channels can be stacked if required by the user). A good prior knowledge of the observed central frequency of the line, i.e. of the redshift of the source, is needed in order to stack the spectra reliably and thus maximize the amplitude of the reconstructed line (see section \ref{res-9}).  
Both median based \citep[similar to][]{Pannella2009} and weighted mean based \citep[similar to][]{Decarli2014} stacking have been developed. Both methods induce a theoretical rms noise reduction by a factor $\sim \sqrt{N}$, where $N$ is the number of sources stacked (if measurements are independent and outliers are symmetric)\footnote{It should be noted that, while rms noise level goes down by a factor $\sim \sqrt{N}$ in both mean and median stacking, the SNR may behave differently when using either. This would be the case if, for example, the studied sample is composed of many dim low-SNR sources, and a few brighter high-SNR sources. The brighter sources would not contribute to the median stack while they would be driving the SNR improvement in the mean stack.}.

\begin{figure*}
\includegraphics[width=\textwidth]{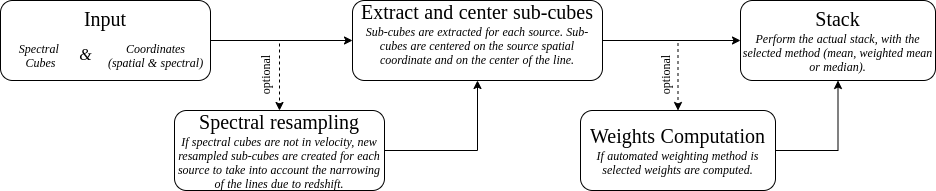}
\centering
\caption{Flow chart showing the main steps taken by the main stacking algorithm of \textsc{LineStacker}.} 
\label{fig:flowChart}
\end{figure*}

Required user inputs to the algorithm are: a list of data cubes, the spatial coordinates of the target sources and the spectral coordinate of the associated line (i.e. either observed central frequency, or redshift of the source and the rest line frequency or central channel index). The spatial and frequency sizes of the subimages are specified by the user prior to stacking. For each target source, data is extracted from its associated cube, and filled into the associated empty subimage. All subimages are then buffered to facilitate access to data, this is especially relevant when using statistical tools implying numerous iterations of stacking. Finally all stamps are stacked together, according to the user specified method (mean, median or weighted-average). If weighted-average stacking is used, weights can be automatically calculated through a set of embedded methods or input by the user. See section \ref{weightings} for a complete description of the automated weighting methods. {The main steps taken by the main stacking algorithm can be seen on Figure \ref{fig:flowChart}.}

It should be noted that a difference between the results from median and mean stacking would imply a skewed distribution of the sources in the studied sample: while mean results would be driven by a few, brighter, outliers, they should have a lesser impact on the median results. Using multiple stacking methods can hence be a good diagnostic of a skewed distribution of the sample. 

See appendix \ref{ex:CubeBasic} and \ref{ex:CubeExtended} for examples using of \textsc{LineStacker}. 

\subsection{Edge treatment in spectra} \label{edge}

When observing it is possible that the emission line of the target source is not centred within the observed spectral window, but falls near the edge. This results in only partial line coverage, and could inhibit the inclusion of such sources in stacking. {In order to still include these sources, for each source, channels outside of the observed window are omitted from the stack. This will result in a certain range of spectral channels in the stack containing less sources}. In such a situation, the noise will have higher values near the edges of the stacked spectrum. When stacking with \textsc{LineStacker}, the user gets, as an additional output, the number of sources used in every spectral bin, in order to take this effect into account when interpreting stacked data.

\subsection{Estimating noise level} \label{noise}

In order to calculate noise level in the data, two methods are available. The first computes the noise on the entire spectrum (through collapsing all frequency channels), while the second handles the noise channel by channel, allowing to account for noise variation with frequency \citep[e.g.][]{Bischetti2018}. In addition, and in both cases, noise levels can be computed either through a user-defined region around the target sources, or across the entire cubes. { Typically, noise levels are used as weights for sources in the stack. Calculating the noise across the entire cubes is therefore more relevant if there is only one source per cube, or at least if the dimensions of the cubes are comparable to the size of the sources. This is, however, left to the user to choose.
Noise is calculated by computing the standard deviation of the data across the selected region.} Computing noise levels across the entire cubes can be useful if, for example, cubes are obtained from different observations, as some may present much higher noise values (due to a lower integration time or varying observing conditions) and should therefore hold a lower weight in the stack. However, doing so leads to the inclusion of pixels far from the center which will hold intrinsically higher noise level due to the reduction of the primary beam response. Computing the noise solely in more compact regions, centered around the target sources, allows to take the source position on the cube into account: sources closer to the phase center should have lower noise and should hence have higher weights.
 
\subsection{Automated weightings} \label{weightings}

If using weighted-average stacking the user can input customized weights for all sources individually or use the automated methods included in \textsc{LineStacker}. Automated methods include:

\begin{itemize}
    
\item Weights inversely proportional to the noise of the cubes: $$W_{\rm i}=\frac{1}{\sigma_{\rm i}^2}$$ with $W_{\rm i}$ the weight of source $i$, and $\sigma_{\rm i}$ the standard deviation associated to source $i$. As stated in section \ref{noise}, noise can be computed across the entire cube. In which case $W_{\rm i}$ is the weight of cube $i$, and $\sigma_{\rm i}$ the standard deviation across cube $i$.
\item  Alternatively different weights can be used for each frequency bin (depending on the individual noise values in that frequency bin), in this case, $W_{\rm i,j}$, weight of source $i$ at spectral channel $j$ is defined as \citep{Fruchter2002}:
$$W_{\rm i,j}= \frac{1}{\sigma^2_{\rm i,j}}$$ where $\sigma^2_{\rm i,j}$ is the standard deviation associated to source $i$ at spectral channel $j$. 

Following this, one can define $W'_{\rm j}$, the total weight on stack channel $j$ as: 
$$W'_{\rm j}=\sum_{\rm i=1}^{\rm n} W_{\rm i,j} = \sum_{\rm i=1}^{\rm n} \frac{1}{\sigma^2_{\rm i,j}}=\frac{1}{\sigma'^2_{\rm j}}$$
where $n$ is the total number of sources and $\sigma'_{\rm j}$ the summed standard deviation at spectral channel $j$. Similarly to above, noise computation can be extended to the entire cube or restricted to regions around the target sources.

\item  If the lines are individually distinguishable before stacking, weighting can be set proportionally to properties of the source, for example: $$W_{\rm i}= \frac{1}{A_{\rm i}}$$ with $W_{\rm i}$ the weight and $A_{\rm i}$ the amplitude of line $i$ \citep[e.g.][]{Stanley2019}.
\end{itemize}

\subsection{Linewidth change in frequency due to redshift: resampling.} \label{regridToZ}

{If the spectral cubes are not sampled in velocity space but in frequency (or wavelength) space instead , the width of the lines emitted by sources at different redshifts will change (getting narrower in frequency with increasing redshifts).} In order to take this effect into account an option is included into \textsc{LineStacker} to resample sub-cubes according to their redshift, before stacking. Two options are available: to either use the line with the highest or lowest redshift as a reference -- using the lowest redshift will lead to over-sampling while using the highest implies under-sampling, see below. The sub-cube associated to the line of reference is kept identical while the other are resampled such that:

$$\Delta \nu_{\rm new}=\frac{\Delta \nu_{\rm old}}{z_{\rm ratio}} \,,$$

which is equivalent to

$$N_{\rm new}=N_{\rm old}*z_{\rm ratio} \,,$$

where $\Delta \nu_{\rm new}$, $\Delta \nu_{\rm old}$, $N_{\rm new}$ and $N_{\rm old}$ are the channel size and number of channels after and before resampling respectively. And $z_{\rm ratio}=\frac{1+z}{1+z_{\rm ref}}$ where $z$ is the redshift of the source being resampled, and $z_{\rm ref}$ is the redshift of the line chosen as reference. Once the new channel size is computed the resampling is performed through linear interpolation. Because such a treatment implies modification of the data -- and may not be relevant if the sources stacked have similar redshifts -- it is optional when using \textsc{LineStacker}. 
{ To avoid over-sampling the data, we would advise against using the smallest redshift as reference.

However, because both methods imply linear interpolation of the data, it is advised to work directly with cubes binned in velocity, when working with a sample with a very large redshift range, thus avoiding this problem from the start.}

\subsection{1D-Stacker} \label{subsect:1DStacker}

In addition to the cube stacker, a module allowing 1D stacking is also included in \textsc{LineStacker}. 
The required user inputs are the spectra and the corresponding line centers. The line centers can also be identified automatically, through Gaussian fitting, if the lines are detectable before stacking. Similarly to cube stacking, individual weights for weighted-average stacking can be input by the user or automatically calculated (see section \ref{weightings} for a detailed description of the different weights). Median stacking is also available. 
\par The 1D module of \textsc{LineStacker} can be used on spectra extracted from cubes beforehand, allowing individual custom spectra extraction for each source. This can be useful if, for example, sources are known to have different spatial extent.  
\par Unlike cube stacking, \textsc{1D-Stacker} does not require CASA functions, and can be run in a Python session. See appendix \ref{ex:1Dextended} for an application example of 1D-Stacker.


\section{Statistical analysis tools} \label{sec:stat}

Here we present the statistical tools included in \textsc{LineStacker} to assist with the analysis of the stack spectra. Some tools are meant to be applied after stacking (post-stacking), to determine the robustness of the stack. Other tools can be used before stacking (pre-stacking) to get a better insight of the distribution of the stacked population. 

\subsection{Estimating the significance of the stack result} \label{stackEmptyPos}

In order to estimate the significance of the stack result one can stack source-free positions and compare the result to the initial stack. For every source, a random position on the map excluding the region around the source is chosen to be stacked. The new set of source-free targets is then stacked, similarly to the original target sources (same weighting scheme etc.).  
This process is performed a large number of times (user defined, typically of order $10^{4}$ - $10^{5}$), as a Monte Carlo process, to reach a good statistical significance of the empty positions (homogeneously probing the field and thus avoiding peculiar or peak noise values). Comparing the distribution of the results from the source-free stacks to the result of the source stack, allows for a good estimate of the significance of the stack. This method shows to what extent the result obtained from stacking the original target sources could be reproduced by stacking only noise.

\subsection{Bootstrapping} \label{bootstrap}

When coupled to stacking bootstrapping can be used to probe the distribution of the parameters of the lines (amplitude of the stacked line, width of the line or integrated flux) in the original sample.

In statistics, bootstrapping methods are methods of statistical inference that allow estimation of the distribution of the sample parameters, through randomly resampling the original sample with replacement. Each source added to the new sample is randomly chosen from the entire pool of sources, allowing for multiple selections of the same source. The total number of possible combinations of resampling N elements is $\Gamma^{N}_{N}=\frac{(2N-1)!}{N!(N-1)!} $ which, for $N=30$, is of order $10^{16}$. Therefore bootstrapping methods are most commonly combined with stochastic methods such as Monte Carlo analysis. See appendix \ref{ex:1Dextended} for an example use of bootstrapping. 

\subsection{Subsampling} \label{subsample}

Subsampling consist in choosing a new, smaller sample of sources from the original target sources. This method is performed by randomly choosing a new sample size (between 1 and $N$, where $N$ is the total number of sources), and randomly filling it with any of the target sources (without replacement\footnote{Once picked the source is removed from the pool, preventing sources to be placed twice in the same sample.}). The stack is then performed again, using this new set of sources. A grade is assigned to sources present in the subsample, depending on how well their stack compares to the original/full stack (the grading system depends on what specific characteristics the user is trying to probe, and hence what kind of test is applied to the data set). Performing this procedure a high number of times allows to identify if some specific subset of sources exhibit an average higher grade. Similarly to bootstrapping, this aims at studying the sample's distribution, but subsampling could allow individual identification of outliers. A good example of the use of subsampling can be found in \citet{Stanley2019}, where we used it to identify sources more likely to show an outflow component out of a sample of high-redshift quasars. See appendix \ref{ex:1Dextended} for an example use of subsampling.

\subsection{Spectral rebinning} \label{rebin}

Spectral rebinning consists in changing the size of the spectral bins of each cube/spectra depending on the width of the line. As shown in section \ref{res-randWidth} stacking lines with different widths will impact the stacked line profile: even if all lines have Gaussian profiles initially, the resulting stacked line will not be Gaussian. This can be a source of bias if trying to give a diagnostic of lines profile (e.g. while looking for outflow signatures). If the linewidth is identifiable pre-stacking, it is possible to change the bins size, individually for each source, so that all lines span the same number of spectral channels. The resulting stacked line will then retain a Gaussian shape. It should be noted that, after such treatment, the channel size of each spectrum can be defined as in \citet{Stanley2019}:
$cw_{\rm rebin}=cw_{\rm origin}\,\times\,$FWHM$_{\rm origin}$/FWHM$_{\rm min}$, where $cw_{\rm rebin}$ is the channel width after rebinning, $cw_{\rm origin}$ is the original channel width, FWHM$_{\rm origin}$ is the full width half maximum (FWHM) of the line before rebinning, and FWHM$_{\rm min}$ is the FWHM of the narrowest line (used as a reference to rebin all the spectra).
After stacking the channel width of the stacked spectrum can be thought of as the mean channel width of all rebinned spectra.
See \citet{Decarli2018QSO,Stanley2019} for example use of spectral rebinning. See appendix \ref{ex:CubeExtended} for an example use of spectral rebinning. 

\section{Simulations} \label{simu}

To evaluate the performances of \textsc{LineStacker} on different observable cases, we simulate data sets mimicking interferometric data. We concentrate on two different data types: the full spectral cubes (3D data cubes), and extracted spectra (1D spectra).
While stacking simulated 3D data sets allows us to characterize the general performances of \textsc{LineStacker}'s main algorithm, stacking 1D spectra permits the study of the effect of complex line profiles on the stack.  
Every set of simulations and the associated analysis is performed 100 times in the case of the 3D data sets and 1000 times for the 1D data sets to increase statistical significance. While multiple weighting schemes are available in \textsc{LineStacker} all data sets are stacked using a weighting of $w=1$ for all sources.
Characteristics of the simulation sets are given in Tables \ref{table:charac3D} and \ref{table:charac1D}. Throughout we assume $H_0=70$km\,s$^{-1}$\,Mpc$^{-1}$, $\Omega_M=0.3$, $\Omega_\Lambda=0.7$.

\subsection{3D simulated data sets: general characteristics} \label{simu3D}
All 3D simulated data sets but two are generated using the CASA task SIMALMA (a task performing simulations of ALMA observations\footnote{\textit{https://casa.nrao.edu/docs/TaskRef/simalma-task.html}}) simulating ALMA cycle 6, configuration C43-2 -- a short baseline configuration: max baseline=314m, corresponding to an angular resolution $\sim 1''$ at 230GHz\footnote{\textit{https://almascience.nrao.edu/tools/proposing/proposers-guide\#section-53}} and a primary beam FWHM of $\sim22''$. This ALMA configuration has been chosen because most studied cases focus primarily on the frequency signature of sources and less on the spatial distribution. We hence favor faster computing time over better spatial resolution. If not specified otherwise, we simulate band 6 observations: with a central frequency of 230GHz and a bandwidth of 4GHz. Unless specified otherwise, all simulations have a velocity resolution of 100\,km\,s$^{-1}$ which corresponds to a frequency resolution of $\sim$ 80MHz.
Furthermore, we simulate VLA observations, using CASA task SIMOBSERVE (a more general simulation task in CASA), simulating configuration C, corresponding to an angular resolution of $0.95''$ and a FOV of $120''$. The central frequency is set to 22 GHz, falling in the middle of the K band. The total bandwidth is  500 MHz with a resolution of 100\,km\,s$^{-1}$, which corresponds to a frequency resolution of $\sim$ 7.3MHz.
We set the total integration time to 20min for each pointing, corresponding to typical ALMA/VLA pointing time. {When the channel width is set in frequency instead of velocity (data set7 and data set8b), we chose to divide the bandwidth in 60 channels, meaning that the average channel width in frequency may slightly differ from the constant 100\,km\,s$^{-1}$ of the other sets (see Table \ref{table:charac3D})}.  

Every source's image is generated through a component list (a list of functional representation of the sky brightness) which serves as a skymodel (model image of the observing field) whose observation is then simulated with the CASA task SIMALMA (or SIMOBSERVE in the case of VLA data). From this simulation we get the visibilities (i.e. the interferometric data), which are then imaged using the CASA task CLEAN\footnote{the CLEAN algorithm was originally described in \citet{HogbomClean}}, the images being the final product that will be stacked. 

In the absence of bright foreground sources the number of CLEAN iterations is set to 0 (i.e. solely imaging), otherwise CLEAN is performed down to a given threshold ($\sim$ 10 times fainter than the bright foreground source). For every source the spectrum consists of an emission line with noise, the noise being directly generated through the SIMALMA or SIMOBSERVE task depending on the data set. 

Every pointing position is selected randomly within a circle of 10 arcmin radius, centered at J2000 3h31m00.00 -27d40m00.00.\footnote{corresponding to the Extended Chandra Deep Field South field, ECDFS \citep{CDFS_presentation} it has been chosen arbitrarily and does not have any impact on the produced data and its analysis.} 

Every sky model consists of 30 sources, distributed on either 30 or 10 images (either 1 source per image in the center, or 3 sources at random positions, see Table \ref{table:charac3D}). The sample size was chosen such that it is small enough to be representative of most stack cases and large enough to present a relevant noise reduction\footnote{when stacking, noise goes down as $\sqrt{N}$ where N is the number of stack positions ; 30 sources corresponds to an average noise reduction of $\sqrt{30} \sim 5.5$ (see section \ref{disc:sourcesNumber} and figure \ref{fig:sqrtN})}. When only one source is simulated in the middle of each image, the data cubes produced have a radius of $6''$. Such a field size suffices because we are not interested in edge effects (since the sources are in the middle of the field). Otherwise, when multiple sources are generated on one image, sources are randomly distributed within the primary beam size of the simulated data ($28''$). This allows to test for effects that may arise with sources far from the center. In both cases the angular resolution is set to $0.25''$ per pixel, the synthesized beam having a size of $\sim 1''$. 

We create 15x2 different simulated data sets: 15 data sets, with specific characteristics, are tested in both a high and a low SNR. 
Each version serves a different purpose. The high SNR ($\approx 200$) sets allow for a near-ideal test of the algorithm, where noise is almost negligible, and shows the best result one can expect. Low SNR ($\approx 1$) sets test the reduction of noise through stacking, but also the limitations when applied to a case where noise levels are important. We mostly concentrate on point like sources but we also examine cases with extended sources. For flux conservation near the edges of the map every simulation has been primary beam corrected. We note that primary corrections should be done by the user as it is not a part of the image stacking routine of \textsc{LineStacker}.

\subsection{Spectra} \label{simuSpectra}

For our simulated data sets we assume a sample of high-redshift galaxies. We chose two scenarios: [C\,\textsc{ii}] emission lines, at $z \sim 7$, observed with ALMA in band 6 and CO(3-2) at $z \sim 4$ observed with VLA in band 4. In both cases the spectra consist solely of the emission line and noise. These lines have been chosen arbitrarily, as typically observed emission lines at high-redshift. The lines and redshift choice should have no impact on the produced data and hence on the conclusions drawn in this article.
The high SNR samples are generated simulating emission lines peaking at $100 \,$mJy, versus $0.5 \,$mJy for the low SNR samples ; $100\,$mJy is typically brighter than what would be expected, this value is picked in order to have near-ideal, very high SNR values. 

In some sets (see Table \ref{table:charac3D}) the amplitude is not fixed, but randomized uniformly over a given range (i.e. top hat distribution). The ranges: from $50\,$mJy to $150\,$mJy and from $0.1\,$mJy to $0.9\,$mJy for the high and the low SNR sets respectively, are chosen such that the mean amplitude reaches the same value as the fixed one, to allow easier comparison between sets. When fixed, the line FWHM is set to 400\,km\,s$^{-1}$ \citep[which is typicall for high redshift galaxies, e.g.][]{Bothwell2013,Gullberg2015,Gullberg2018,Decarli2018QSO}. When random, the FWHM is randomized uniformly from 200\,km\,s$^{-1}$ to 1000\,km\,s$^{-1}$. Exact values do not have a big impact and are chosen to be representative of observed values where large variations in line widths are seen \citep[e.g. ][]{Bothwell2013,Decarli2018QSO}. Throughout the text line width will refer to the FWHM of the line.

\subsection{Simulated data sets: 3D} 

In Table \ref{table:charac3D} we give the characteristics of each 3D simulated data set, and here we provide further information on each data-set.

\subsubsection{Most basic case - set1a} \label{simu-mostbasiclowres}

To test the performance of \textsc{LineStacker} in a near-ideal, fully controlled simulation, we start with the simplest [C\,\textsc{ii}] line simulation, using a realistic spectral resolution. 
It consists of a point source in the center of every image, with all the same spectrum which is simply a [C\,\textsc{ii}] emission line with a width of 400\,km\,s$^{-1}$
A total of 30 cubes are simulated, with a channel width of 100\,km\,s$^{-1}$, which is typical of stacked data where the channels are collapsed together beforehand in order to maximize the amplitude of the signal to the detriment of frequency resolution. This set will be considered as a reference, to be compared to the following sets to see the impact of the tested parameters.

\begin{table*}[htbp]
\caption{Characteristics of 3D simulated data sets.}
\centering
\resizebox{\textwidth}{!}{\begin{tabular}{ccccccccccccc}
  \hline
  Data set$^{(a)}$&\# sources$^{(b)}$& \# cubes$^{(c)}$&{Line} peak$^{(d)}$&Linewidth$^{(e)}$&velocity$^{(f)}$ &Position$^{(g)}$&FOV$^{(h)}$&Source size$^{(i)}$&Simulation$^{(j)}$& Redshift$^{(k)}$ & Foreground$^{(l)}$&\\
   &per cube& &randomization & FWHM &resolution & &  &  &type&range&source&\\
   & & &&(\,km\,s$^{-1}$)&(\,km\,s$^{-1}$)&  &(arcsec)&  &\\
  \hline

  1a & 1 & 30  &  No & 400 & 100  & center & 6 & PS & ALMA & $7.19<z<7.32$ & No \\ 
  
  1b & 1 & 30 &  No & 400 & 10  & center  & 6 & PS & ALMA & $7.19<z<7.32$ &  No \\

   2a & 1 & 30 &  Yes & 400 & 100  & center & 6 & PS  & ALMA & $7.19<z<7.32$ & No \\
   
   2b & 3 & 10 &  Yes & 400 & 100  & random & 28 & PS & ALMA & $7.19<z<7.32$ & No \\

   3 & 1 & 30 &  No & 200-1000 & 100  & center & 6 & PS & ALMA & $7.19<z<7.32$ & No  \\

   4 & 3 & 10 & Yes & 200-1000 & 100  & random & 28 & PS & ALMA & $7.19<z<7.32$ & No \\
  
   5a, 5b, 5c & 1 & 30 & No & 400 & 100  & offset & 28 & PS & ALMA & $7.19<z<7.32$ & Yes  \\
   
   6a & 1 & 30 & No & 400 & 100  & center  & 6 & $0.1''$ - $1.5''$& ALMA &  $7.19<z<7.32$ & No \\
   
   6b & 1 & 30 & No & 400 & 100  & offset & 6 & $0.1''$ - $1.5''$ & ALMA & $7.19<z<7.32$ & No  \\
   
   7 & 3 & 10 & Yes & 400 & $\sim80$  & random  & 28 & PS & ALMA & $5.89<z<7.99$ & No \\
   
   8a & 3 & 10 & Yes & 200-1000 & 100  & random  & 120 & PS & VLA & $4.12<z<4.36$ & Yes \\
   
   8b & 3 & 10 & Yes & 200-1000 & $\sim120$  & random  & 120 & PS & VLA & $3.34<z<5.40$ & Yes \\
   
   9 & 1 & 30 & No & 400 & 100  & center  & 6 & PS & ALMA & $7.19<z<7.32$ & No \\
   
\hline

\end{tabular}
\label{table:charac3D}

}
\begin{flushleft}

$(a)$ Data set number ID; $(b)$ Number of stacking target sources on each cube; $(c)$ Total number of cubes stacked for each stack iteration; $(d)$ Is the [CII] peak value randomized ("Yes") or fixed ("No"); $(e)$ FWHM of the simulated Gaussian emission lines; $(f)$ Velocity resolution (i.e. size of the spectral channels) of the simulated cubes; $(g)$ Spatial position of the sources on the cubes: on the center, randomized or offset.; $(h)$ Total spatial extent of the simulated cube; {$(i)$ Physical source size, "PS" stands for point source. Corresponding sizes once convolved with ALMA beam are $\sim 1''$ for the point sources (and $\sim 0.95''$ in the VLA case), from $\sim 1''$ to $\sim 1.8''$ in the case of extended sources;} $(j)$ Interferometer simulated for the observations. {$(k)$ Redshift range of the target sources. $(l)$ Presence of bright foreground sources. }
\end{flushleft}

\end{table*}

\begin{table*}
\caption{Characteristics of 1D simulated data sets.}
\centering
{
\begin{tabular}{cccccccccc}
  \hline
  Data set$^{(a)}$&Number of Sources$^{(b)}$&Linewidth$^{(c)}$&velocity resolution$^{(d)}$&spectral signature$^{(e)}$& $\Delta z$$^{(f)}$ \\
   & &(\,km\,s$^{-1}$) & (\,km\,s$^{-1}$)& & \\
  \hline

   10 & 30 & 400 & 100  & Gaussian & True \\
  
   11a & 30 & 100-700 & 100 & double peaked & False \\
   
   11b & 30 & 400 & 100 & double peaked & True \\
  
   12 & 30 & 400 & 100 & two Gaussian components & False \\
   
   13 & 30 & 400 & 100  & Gaussian & False \\
\hline
  
\end{tabular}
\label{table:charac1D}

}
\begin{flushleft}
 $(a)$ Data set number ID; $(b)$ Total number of stacked spectra for each stacking iteration; $(c)$ FWHM of the simulated emission lines; $(d)$ Velocity resolution (i.e. size of the spectral channels) of the simulated spectra; $(e)$ Spectral shape of the simulated emission lines; $(f)$ Usage, or not, of uncertainties on the observed redshift, resulting in uncertainties in the central line position.
 \end{flushleft}
\end{table*}

\subsubsection{Most basic case, high spectral resolution - set1b} \label{simu-mostbasichighres}

When the emitted line is observed, it is binned to a finite number of channels. When the number of channels across the line is small (of order $\sim 5$ across the line FWHM) the line's amplitude will be systematically under-evaluated. To test this effect, and show that the amplitude loss is due to the finite spectral resolution and not to some other intrinsic bias, we repeat data set1a with a much better spectral resolution. The number of channels is 10 times higher than in set1a for the same bandwidth. Even though such a spectral resolution is better than the one typically expected, the goal here is to build a near-perfect reference set, estimating the signal loss that can be expected due to the finite number of channels.

\subsubsection{Random amplitude - set2a} \label{simu-randomAmpA}

Data set2a is designed to check the effect of a given distribution of the lines amplitude on the stack. Amplitudes of each line are randomized uniformly on a range of $50\,$mJy to $150\,$mJy and $0.1\,$mJy to $0.9\,$mJy for the high and low SNR sets respectively. The width of every line is kept at 400\,km\,s$^{-1}$.

{
\subsubsection{Random amplitude and position - set2b} \label{simu-randomAmpB}

{
Stacking can be expected to be used on individual cubes containing multiple sources. Because \textsc{LineStacker} performs stacking on sub-cubes the effect coming from such configuration should be limited. However such a case of configuration is likely to be a common user case, and therefore is relevant to test.}
Here, unlike set2a the number of sources per image is set to 3, with positions uniformly randomized across the whole field. Amplitudes of each line are still randomized uniformly on a range of $50\,$mJy to $150\,$mJy and $0.1\,$mJy to $0.9\,$mJy for the high and low SNR sets respectively. The width of every line is kept at 400\,km\,s$^{-1}$.}

\subsubsection{Random linewidth - set3} \label{simu-randomWidth}

Data set3 aims at quantifying the effect on the stacked line when including lines with a range of line widths. This situation is expected in real data since galaxies exhibit a range of line width due to different masses, orientation and level of turbulence.
In order to properly evaluate the consequence of such an inhomogeneous distribution this data set will be based on set1a but with a FWHM randomly picked from a uniform top-hat distribution from 200\,km\,s$^{-1}$ to 1000\,km\,s$^{-1}$.

\subsubsection{Random amplitude, linewidth and position - set4} \label{simu-allofit}

Data set4 is a combination of data set2b and 3, with randomized positions on the field as well as randomized amplitude and width of the line. 

\subsubsection{Bright foreground source in the center - {sets 5a, 5b and 5c}} \label{simu-bright}

These data sets aim at quantifying the effect that could arise from stacking faint sources from a map containing a bright central point source. Such a case can be expected when stacking faint peripheral sources present in a field centered on a bright source. The presence of bright sources (continuum or spectral line) affects the quality of the interferometric image products, as imperfect modelling of the bright source can leave artefacts in the final data cubes; such artefacts increase the noise and could potentially affect the stacking result. 
In the simulations, the central bright source has a continuum flux density of 1\,Jy and 0.1\,Jy for high and low SNR simulations respectively.
While 1\,Jy is higher than typically expected values, the foreground source has to be brighter than the amplitude of the line of the target sources -- which are already very bright in the high SNR sets. The target sources have the same properties as sources from data set1a. { To properly diagnose the impact of the bright foreground source we created three type of data sets. In data set5a the target source is located at a distance of $2.5''$ from the bright source, $5''$ in data set5b and $10''$ in data set5c.}
The bright foreground source was removed using the CASA task CLEAN, and the stacking was performed on the residual image.

\subsubsection{Extended sources - set6a} \label{simu-extendedA}

Data set6a is composed of extended sources, to test cases where sources are resolved.
We investigated such a case by simulating a data set similar to data set1a, but containing extended sources. Sources have Gaussian shape, with their orientation and size taken at random. The length of the major and minor axes are randomized uniformly between $0.1''$ and $1.5''$ {(this is the physical extent of the sources and corresponds to a size range of $\sim 1.00''$ to $1.80''$ after convolution with the ALMA beam).} The orientation and size of the sources will have an important impact since different sources will not be spanning the same area. Each source is simulated with a spatially integrated peak flux value of 100\,mJy and 0.5\,mJy for high and low SNR sets respectively.

\subsubsection{Extended sources with an offset - set6b} \label{simu-extendedB}

Typically, the stacking positions of sources are defined based on observations at other wavelength. This can induce an offset between stacking position and source position at the observed wavelength. In the case of the observation of emission lines the region of interest may be offset from the position defined at a different wavelength (for example, [C\,\textsc{ii}] may be distributed differently on the studied galaxy than dust or the stellar population e.g. \citealt{Rybak2019,Fujimoto2019}). In order to reproduce and quantify such an effect, we produced another data set, similar to data set6a, but where the stacking position is offset compared to the source position. This offset is taken at random within a uniform distribution between $0''$ and $1''$. This will impact the stack results since sources will not be properly aligned when stacked. {It should be noted that point sources are of course also subject to potential offset. However, due to the nature of the observations, their observed size will be equal to size of the synthesized beam. The effect coming from the offset of point sources is hence comparable to the effect tested in this data set.}

\subsubsection{Random central frequency, larger redshift range - set7} \label{simu-largeBW}

Data set7 is based on data set2b but studies sources spanning a larger redshift range and thus a larger observational central frequency range. The redshifts are drawn randomly from a uniform distribution in the range $5.89 < z < 7.99$ (corresponding to a range of possible observed central frequencies covering the entire band 6). {The bin width is set to a fixed frequency size of $\sim$\,67 MHz, 60 channels over the 4 GHz bandwidth, corresponding to $\sim$\,80\,km\,s$^{-1}$ for the average frequency.} Trying to stack observations done on a large frequency range will have different implications, one of the most direct is the shape and size of the beam, which changes with frequency and could hence be different from one observation to the other. In addition, because of the large redshift range, it is necessary to take into account the  change of width of the lines when working in frequency. 
Each sub-cube is therefore resampled before stacking according to the method described in section \ref{regridToZ}. The highest redshift is used as reference for resampling, leading to a reduced number of channels after resampling.

\subsubsection{VLA type simulation - set8a} \label{simu-VLAa}

We simulate CO(3-2) observations with VLA.
To both showcase that the tool is not solely usable on ALMA type data and also to study cases with a larger field of view and more polluting bright sources. On each cube we simulate 3 target sources with characteristics similar to data set4. Additionally, we add: one very bright foreground point source with a line amplitude of 1\,Jy in both the high and low SNR sets, as well as two ``medium bright" foreground point sources with amplitude 100\,mJy and 10\,mJy in the high and low SNR sets respectively. All sources positions are uniformly randomized across the whole field. The frequency coverage for this set corresponds to a redshift range of $4.12<z<4.36$. {It should be noted that since the line amplitudes of the target sources are kept the same as in the other sets, the SNR will be slightly reduced (due to the lower sensitivity of simulated VLA data): SNR\,$\sim$\,1 in our simulated ALMA data corresponds to SNR\,$\sim$\,0.8 in our simulated VLA data.}

\subsubsection{VLA type simulation, larger redshift range - set8b} \label{simu-VLAb}

Set8b is an extension of set8a to the entire K band, the central frequency is chosen at random so that the entire bandwidth is contained between 18 and 26.5 GHz. {Bandwidth is kept the same as in set8a but channel size is kept constant in frequency at $\sim$\,8.3 MHz: 60 channels over the 500 MHz bandwidth, corresponding to $\sim$\,112\,km\,s$^{-1}$ for the average frequency}. The goal is similar to set7: to study the impact { of the large redshift range on the stack, but with a larger redshift range: from $3.34<z<5.4$. Similarly to set7 a resampling is applied to each sub cube to take into account the redshift of the sources.}

\subsubsection{1D stack of spectra extracted from cube - set9} \label{simu-cubeTo1D}

To allow easy comparison between the 3D and the 1D data sets, we built a data set where the spectra are extracted from individual cubes and then stacked using the 1D module of \textsc{LineStacker}. The sources' properties are the same as set1a and the spectra are extracted from the central pixel.

\subsection{Simulations data sets: 1D} \label{simu1D}

The 1D simulated data sets allow us to test some specific spectral signatures more easily than when using full 3D simulations: we examine cases of double peaked line profiles, outflow signatures, the effect of redshift uncertainties on the stack, and the effect of stacking lines located on the edge of the observed spectral window.
The data sets are generated with a bandwidth of 3000\,km\,s$^{-1}$, and a resolution of {100}\,km\,s$^{-1}$. These spectral only simulations are not generated through CASA, allowing faster computing time. Individual spectra are generated by creating individual Gaussian components and adding randomly generated Gaussian noise on top of each channel. Similarly to the 3D sets two SNR configuration: one with high (pre-stacking) SNR ($\sim 200$) and one with a SNR of order unity. Lines are generated with an amplitude of 200\,mJy in the high SNR data sets and 1\,mJy in the low SNR data sets. The noise follows a gaussian distribution centered at 1\,mJy. Linewidths differ from simulation to simulation, see Table \ref{table:charac1D} for a complete description of the characteristics of each simulated data set.

\subsubsection{Diagnostic of redshift uncertainties - set10} \label{simu-dz}

One of the biggest challenges when stacking lines of distant galaxies arises from redshift uncertainties.
Every simulation performed previously has been computed expecting a perfectly good knowledge of the redshift. But, realistically, redshift is never known with 100\% accuracy, and redshift uncertainties can cause lines not to be stacked on the same central frequency.

Consequently, the amplitude and width of the stacked line will be washed out and potentially become indiscriminate from the noise. In order to quantify this problem we construct spectra datasets, and test different levels of redshift uncertainties ($\Delta z$). The linewidths are all set to a velocity of 400\,km\,s$^{-1}$.
The redshift uncertainties, $\Delta z$, are set to values of 0, 0.001, 0.005, 0.01, 0.05 and 0.1 {(corresponding to velocity shifts of $\sim$ 0, 36, 180, 360, 1800, 3600\,km\,s$^{-1}$)} and the observed redshifts of the sources in each simulated data set are chosen randomly between $z_{\rm true}\pm\Delta z$, where $z_{\rm true}$ is the real redshift of the line. For every $\Delta z$, 1000 data sets are created.

\subsubsection{Double peaked spectrum - set11a} \label{simu-doubleA}

Rotational signatures from galaxies can be seen as double peaked line profile. In order to study such case we create a simulated 1D sample with such properties. The spectra were designed to have a distance, D, between the two peaks ranging uniformly from 200\,km\,s$^{-1}$ to 600\,km\,s$^{-1}$, and an amplitude of 10\,mJy, while the width of the line ranges from 100 to 700\,km\,s$^{-1}$ (with the same linewidth for both components). The central frequency for stacking is taken as the center of the two peaks.

\subsubsection{Double peaked spectrum with redshift uncertainties - set11b} \label{simu-doubleB}

This data set, like the previous one, shows spectra with double peak line profile. This time all lines are simulated with the same characteristics, a width of 400 \,km\,s$^{-1}$ and a distance between the peaks of 400\,km\,s$^{-1}$. An uncertainty on the redshift is added. Like data set10 the $\Delta z$ values of 0, 0.001, 0.005, 0.01, 0.05 and 0.1 are tested. Hence trying to quantify the distortion of this specific spectral signature when confronted to redshift uncertainties, and the extent to which one can recover such spectral characteristics when stacking.

\subsubsection{Outflows - set12} \label{simu-outflow}

In cases where outflows are present, the galaxy's emission lines could include a second, fainter and broader, component \citep[see][]{Stanley2019}. In many cases observation will not be deep enough to detect this signature, and therefore stacking is a useful tool. From a testing perspective, studying outflows allows us to use line stacking in a different fashion: to look for specific spectral signatures below the noise, while the main line is visible: using the main line not as a source but as a reference to stack signal below it. All the lines are simulated with a width of 400\,km\,s$^{-1}$, and a broad component with a width of 1000\,km\,s$^{-1}$. {The amplitude of the broad component is one tenth of the amplitude of the main line. The two SNR configurations (200 and 1) are based on the broad component amplitude. The amplitude of the main lines will be 2\,Jy in the high SNR configuration and 10\,mJy in the low SNR regime -- corresponding to an amplitude of the broad component of 200\,mJy and 1\,mJy in the high and low SNR configuration respectively.}
   
{
\subsubsection{Lines on the edge - set13} \label{simu-linesOnEdge}

One of the strengths of \textsc{LineStacker} is its ability to stack lines located on the edge of the observed spectral window. To showcase this capability and test its performance we produced a data set where all lines are located on the edge. While it is unrealistic to have a data set where all lines are so far from the center, we decided to test such an extreme case to demonstrate the expected result in the worst case scenario. All simulation parameters are similar to data set10 with a redshift uncertainty $\Delta z=0$, but lines are centered at a distance uniformly randomized between 0 and 200\,km\,s$^{-1}$ from either of the spectral edges. This means that a significant part of the line will be outside of the observed spectral window -- in the worst case where the line is centered exactly on the edge, 50\% of the line will not be observed, in the best case, when the line is centered at a distance of $\frac{1}{2}{\rm FWHM}=$200\,km\,s$^{-1}$, roughly 30\% of the line is outside of the observed window.
}


\section{Results} \label{sec:results}

We used \textsc{LineStacker} on the simulated data sets described in section \ref{simu}. 
From our stacking analysis we extract the amplitude and the width of the line as well as the integrated flux and compare them to the mean input values. Stacking is performed with subimages of size $16\times16$ pixels and $32$ spectral channels.
To retrieve the amplitude and width we fit the resulting stacked line with a Gaussian and extract the fit's parameters. The integrated flux is computed by summing the {flux in each channel covering the detected line and multiplying by the channel width}. The spectral area of integration has a size of 2 times the input line FWHM, centered on the line central frequency. When the amplitude and/or width are simulated at random we calculate their average and use this average for comparison with stack values. Presented reconstruction fractions are the ratio { $(1-|1-\frac{measured}{expected}|)\times100$ for each parameter in both low and high SNR configurations (chosen such that the reconstruction rate is always $<100\,\%$)}. 

{The results, average of all the stacks, are presented in Table \ref{table:result}}. Presented standard deviations are the standard deviations of the stack results across the studied simulation set. In the case of 1D simulations, different parameter reconstruction are tested for every simulation, results from stacking analysis of each one dimensional data-sets are shown in Tables \ref{table:deltaZ1}, \ref{table:doublePeak-result} and \ref{table:outflow-result}.

\subsection{Stacking results from 3D simulated data sets} \label{3DResults}

We first present the results from stacking the 3D simulated data sets. For every data set we analyze separately the average results from the two SNR cases and compare them to the average input parameters. The presented results, for a given data set and a given SNR, are average of 100 stacks. Each studied parameters (amplitude, width and integrated flux) are discussed individually.

\subsubsection{Most basic case - set1a } \label{res-1a}

\begin{figure}
\includegraphics[height=60mm]{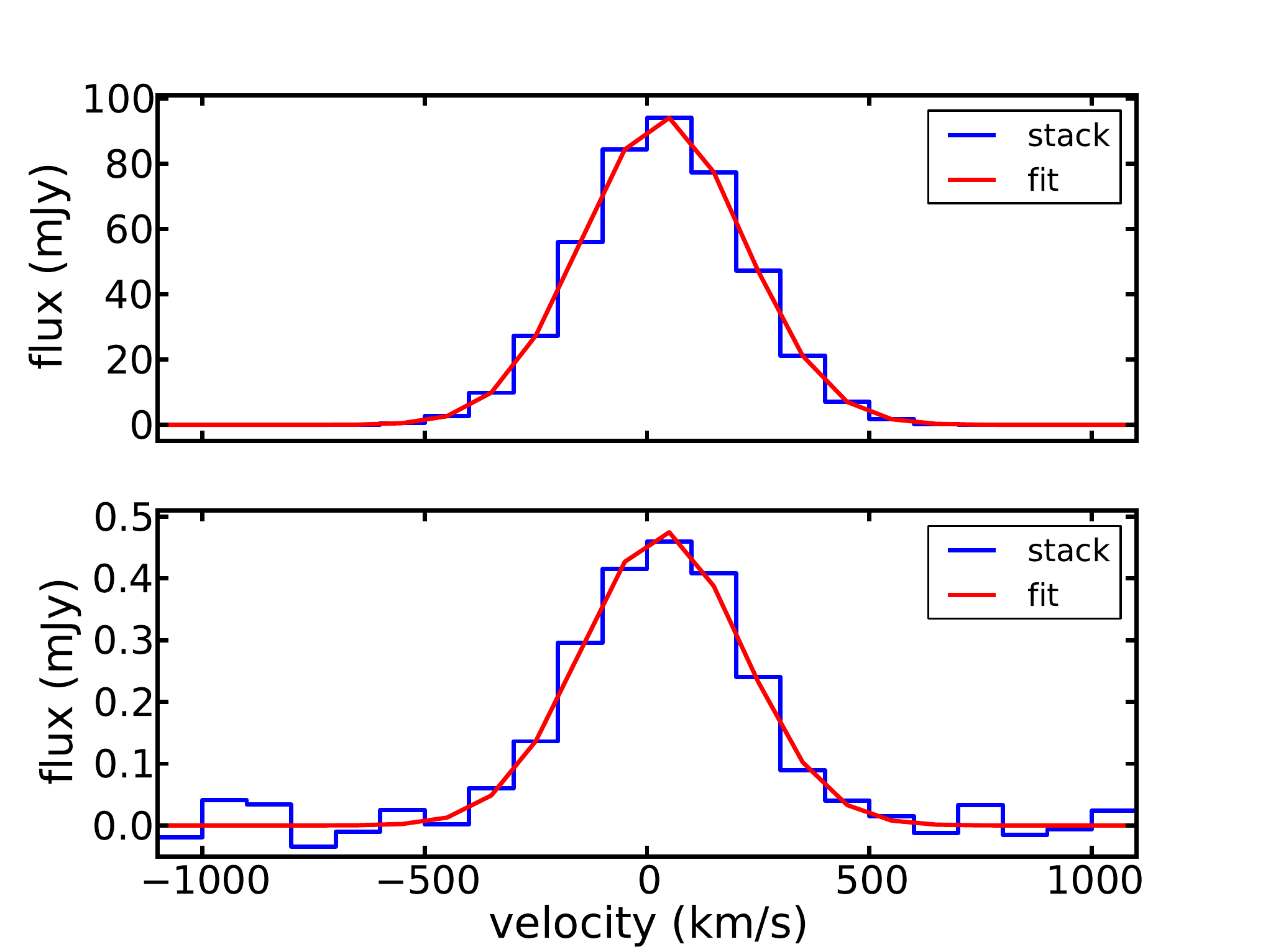}
\centering
\caption{Two example stack of 30 sources from data set1a and the corresponding Gaussian fit.  {\it(top)} high SNR configuration. {\it(bottom)} low SNR configuration. }
\label{fig:dataSet1a}
\end{figure}

After stacking, and fitting our result with a Gaussian, we find a reconstruction fraction of 94.0$\pm 0.22$\%, 92.7$\pm0.23$\% and 99.6$\pm0.16$\% for the amplitude, width and integrated flux of the line in the high SNR configuration. Similar results are found for the low SNR sets: 95.4$\pm5.2$\%, 97.4$\pm5.21$\% and 97.1$\pm2.36$\%. As it will be shown in the next section the missing 6\% in the amplitude reconstruction are systematic errors due to the low velocity resolution. Even though line amplitude is at the same level as the noise in the faint sets, the reconstruction is extremely accurate. One should note that reconstruction of the integrated flux has a stronger dependence to SNR than the amplitude of the line does. This is due to the fact that, to have a proper integrated flux reconstruction, one will need proper reconstruction in every bin containing the line, meaning also the channels containing the outer part of the line profile, which, if the SNR is low, will be under the noise level. {The reconstruction rates obtained from this data set will be used as references when rating success of following data sets, as data set1a has been designed to be the simplest data set, and will hence yield the best results.} Example results from stacking data set1a can be seen on Figure \ref{fig:dataSet1a}. 

\subsubsection{Most basic case, high frequency resolution - set1b} \label{res-1b}

This data set aimed at simulating a near-perfect configuration, differing from set1a with a velocity resolution of 10\,km\,s$^{-1}$, 10 times higher than previously. As expected we reach reconstruction fractions of 99.9$\pm0.02$\%, 99.0$\pm0.06$\% and 100.0$\pm0.05$\%
and 99.4$\pm3.8$\%, 97.7$\pm5.23$\% and 96.5$\pm4.72$\% for the amplitude, width and integrated flux of the line compared to the input of the high and low SNR sets respectively. Showing, as mentioned in the previous section, that the missing 6\%, when reconstructing the amplitude of the line in the set1a, were systematic errors due to the channels width. The almost perfect reconstruction in the low SNR case shows that random errors should be negligible in our stacking setup.

\subsubsection{Random amplitude - set2a} \label{res-randAmpA}

This set has a uniform distribution of amplitude between $50\,$mJy to $150\,$mJy and $0.1\,$mJy to $0.9\,$mJy in the high and low SNR sets respectively.
Here again we find a good reconstruction fraction of 94.0$\pm4.66$\%, 92.7$\pm0.24$\% and 99.6$\pm4.92$\% of the amplitude, width and integrated flux in the high SNR case, and 95.9$\pm9.24$\%, 97.2$\pm5.74$\% and 96.7$\pm8.02$\% in the low SNR case. 

\subsubsection{Random amplitude and position - set2b} \label{res-randAmpB}

This set has a uniform distribution of amplitude between $50\,$mJy to $150\,$mJy and $0.1\,$mJy to $0.9\,$mJy in the high and low SNR sets respectively, with 3 sources per image at random positions.
Here again we find a good reconstruction fraction of 94.1$\pm5.72$\%, 92.6$\pm1.15$\% and 99.7$\pm6.12$\% of the amplitude, width and integrated flux in the high SNR case, and 94.4$\pm8.4$\%, 92.7$\pm8.09$\% and 99.1$\pm8.45$\% in the low SNR case respectively. While flux loss could have been expected with sources far from the pointing center, it is not observed here due to the use of primary beam correction.

\subsubsection{Random width - set3} \label{res-randWidth}

\begin{figure}
\includegraphics[height=60mm]{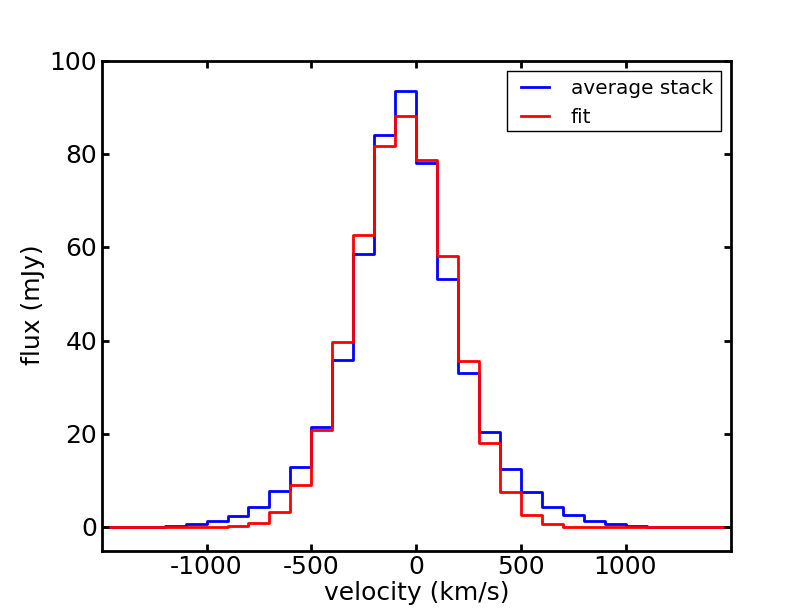}
\centering
\caption{Average stacked line from the 100 realizations of set3, high SNR. Here it is clear -- by comparing with the fit -- that, even if all input lines are Gaussian, the output stacked line is not anymore. One direct consequence of such behavior is the under-evaluation of the amplitude of the line, if done through fitting. }
\label{fig:badFit}
\end{figure}

Here amplitudes and positions are fixed. The linewidth however is randomized, drawn from a flat distribution, between 200\,km\,s$^{-1}$ and 1000\,km\,s$^{-1}$.
When stacking Gaussian lines with different linewidth the resulting stacked line will not conserve a Gaussian shape, and can introduce what could be interpreted as a second component to the line. Fitting the stacked line with a Gaussian to extract parameters will hence give slightly biased results. As shown on Table \ref{table:result} the amplitude is still well reconstructed -- at 90.2$\pm1.59$\% and 92.4$\pm4.0$\% for the high and low SNR simulation sets respectively. Such a difference with set1a is due to the use of Gaussian fitting to extract parameters. As a result the fitted amplitude is lower that expected (see Figure \ref{fig:badFit}). It is, however, interesting to note that, if, instead of fitting, we look at the maximum bin value, it shows $\sim$96\% reconstruction in both SNR cases(see section \ref{disc:underEval}). The width and mean integrated flux are well reconstructed, with a 92.4$\pm8.08$\% and 95.7$\pm8.22$\% reconstruction for the high SNR case, and 99.7$\pm8.49$\% and 91.7$\pm7.19$\% for the low SNR one respectively.

\subsubsection{Random amplitude, width and position - set4} \label{res-allOfIt}

This set includes random amplitudes, random positions, and random width, it is a combination of data sets 2 and 3 and has similar results. Similarly to set3 the reconstructed amplitude is only reconstructed at
90.2$\pm5.32$\% and 90.4$\pm7.6$\% in the high and low SNR setup. This is again due to the intrinsic bad fit, due to a non Gaussian shaped stack line. Width and integrated flux are reconstructed at 91.8$\pm7.0$\% and 96.2$\pm9.18$\% in the high SNR case, and 92.8$\pm11.47$\% and 95.2$\pm10.44$\% in the low SNR one.

\subsubsection{Bright foreground source in the center - sets 5a, 5b and 5c} \label{res-bright}

These sets aimed at studying cases where target sources are faint {and at a fixed distance from a central bright foreground continuum source}. Sources' characteristics are similar to set1a (but target sources are not in the cube center). The amplitude, width and integrated flux are reconstructed at 92.8$\pm0.31$\%, 92.4$\pm0.22$\% and 99.9$\pm0.19$\% respectively, in the high SNR case, and 92.8$\pm6.2$\%, 94.2$\pm5.12$\%, 97.6$\pm8.02$\% respectively in the low SNR one for set5a. Set 5b shows reconstruction rates of 92.4$\pm0.27$\%, 92.4$\pm0.29$\% and 99.4$\pm0.15$\% for the amplitude, width and integrated flux respectively in the high SNR case and 93.4$\pm4.8$\%, 91.6$\pm7.71$\% and 99.5$\pm5.66$\% respectively, in the low SNR one. Finally, set 5c shows reconstruction rates of 92.0$\pm0.27$\%, 92.4$\pm0.23$\% and 99.0$\pm0.23$\% for the amplitude, width and integrated flux respectively in the high SNR case and 93.6$\pm5.0$\%, 94.8$\pm7.28$\% and 97.2$\pm7.55$\% respectively in the low SNR one. The presence of the bright source is well handled through CLEANING and seems to have close almost no impact on the stack.  

\subsubsection{Extended sources - set6a} \label{res-extendedA}

For the cases where sources are extended (set 6a and 6b) the stacked sources are fitted with a 2 dimensional Gaussian and the stacked spectra are extracted from a circular region of radius one FWHM, centered on the stacked source. Fluxes values are converted from Jy/beam to Jy/pixel. This set shows good reconstruction fractions: 95.29 $\pm2.88$\%, 92.85 $\pm0.25$\% and 96.96$\pm2.94$\% in the high SNR case and 90.2$\pm10.2$\%, 90.58$\pm7.64$\% and 79.72$\pm6.6$\% in the low SNR case, for amplitude, width and integrated flux respectively. Indicating that, if the stacking position is well known, stacking extended sources should yield similar results as stacking point sources. It should be noted however, in the low SNR case, the line width is underestimated, leading to an even worse estimate of the integrated flux. This issue arises in extended sources because the outer pixels of the source have a lower line amplitude (by construction), leading to a worse reconstruction of the associated spectra. {In addition, the region from which the spectra are extracted can be more easily underestimated in the low SNR case. Consequently, some of the extended emission will not be successfully retrieved}.

\subsubsection{Extended sources with an offset - set6b} \label{res-extendedB}

Set 6b is build similarly to set6a, but stacking positions are off by a random factor, drawn uniformly between 0 and $1''$. Similarly to set 6a the stacked sources are fitted with a 2 dimensional Gaussian and the stacked spectra are extracted from a circular region of radius one FWHM, centered on the stacked source. As expected, amplitude reconstruction, as well as integrated flux, are not as good as in set 6a, and this effect is even more pronounced in the low SNR configuration. The amplitude, width and flux being recovered at 91.11$\pm2.72$\%, 92.85$\pm0.28$ and 92.60$\pm2.72$ in the high SNR case and 81.6$\pm10.4$, 83.68$\pm8.92$ and 66.04$\pm6.13$ in the low SNR case. If the stacking positions were off by a too important factor then the line reconstruction would eventually be impossible. It should however be noted that, while the reconstruction fraction is inversely proportional to the uncertainties on the stacking positions, it is also proportional to the source size. Hence, the effect coming from position uncertainty will be mitigated by the extent of the sources.  

\subsubsection{Random central frequency, larger redshift range - set7} \label{res-largeBW}

This set is similar to set2b but lines are simulated over a larger redshift range. Bins are set to a fixed frequency size and rebinned to take into account the line width change due to redshift. Amplitude, linewidth and integrated flux are properly reconstructed in both the high and low SNR simulations: 95.9$\pm4.8$\%, 96.1$\pm0.42$\% and 99.5$\pm4.93$\% for the amplitude, width and integrated flux, respectively in the high SNR case and 97.8$\pm10.26$\%, 97.3$\pm8.08$\% and 99.1$\pm11.57$\%, respectively in the low SNR one. {The reconstruction rate for the amplitude and flux is slightly better than in set 2b due to the overall smaller channel width (see Table \ref{table:charac3D}). The standard deviation is, however, higher due to the channel width being fixed in frequency, and hence varying in velocity space.} The near perfect reconstruction shows that we properly correct the effect coming from large redshift range.

\subsubsection{VLA type simulation - set8a} \label{res-VLAa}

In this data set we simulate VLA observations. The reconstruction fractions are 91.1$\pm4.36$\%, 93.8$\pm7.75$\% and 99.5$\pm8.78$\% for the amplitude, width and integrated flux respectively in the high SNR case, and 86.4$\pm17.16$\%, 96.9$\pm14.29$\% and 87.6$\pm12.12$\% respectively in the low SNR one. While parameters are well retrieved in the high SNR case, a lower  reconstruction fraction is observed in the low SNR case compared to ALMA simulations. This is due to a worse sensitivity of our VLA simulations, as mentioned in section \ref{simu-VLAa}.

\subsubsection{VLA type simulation, larger redshift range - set8b} \label{res-VLAb}

 Here we are looking for potential effects that could come from the large redshift range such as size difference in the primary beams as well as side effects from our resampling method, needed to stack sources with large redshift difference and not binned in velocity. Similarly to set7, a larger redshift range does not seem to have any substantial effect on the stacked data. While the reconstruction fractions are slightly lower than the ones in set8a, this is simply due to the velocity resolution which is $\sim$\,20\% higher (see Table \ref{table:charac3D}). Reconstruction fractions are 90.0$\pm4.75$\%, 93.5$\pm8.16$\% and 98.7$\pm9.92$\% for the amplitude, width and integrated in the high SNR case and 82.7$\pm12.27$\%, 92.8$\pm18.27$\% and 88.8$\pm13.66$\% in the low SNR one.

{
\subsubsection{Spectra extracted from cube - set9} \label{res-cubeTo1D}

This data set is similar to set 1a but the spectra are individually extracted from each cube and stacked using the 1D module of \textsc{LineStacker}. The reconstruction fractions are very similar to the one in set1a: 94.0$\pm0.21$\%, 92.7$\pm0.23$\% and 99.3$\pm0.16$\% for the amplitude, width and integrated in the high SNR case and 95.0$\pm4.2$\%, 97.4$\pm4.5$\% and 96.7$\pm2.83$\% in the low SNR one. Which shows a good agreement between our two stacking methods, and justifies our usage of 1D data sets as an easier way to test specific spectral effects in stacking. 
}

\begin{table*}
\caption{Stacking results from all 3D simulations. Presented results are obtained by averaging 100 stacks. Amplitude and linewidth are obtained through Gaussian fitting, while integrated flux is obtained by integrating a given number of channels (see section \ref{sec:results}). Presented errors are computed standard deviation of the given parameter in the 100 stacks.}
\centering

{
\begin{tabular}{cccccccccc}
  \hline
  Data set$^{(a)}$&
  \multicolumn{3}{c}{Mean line amplitude$^{(b)}$} &
  \multicolumn{3}{c}{ Mean linewidth$^{(c)}$} &
  \multicolumn{3}{c}{ Mean integrated flux$^{(d)}$}\\
  &
  \multicolumn{3}{c}{(mJy)} &
  \multicolumn{3}{c}{ (\,km\,s$^{-1}$)} &
  \multicolumn{3}{c}{ (Jy\,km\,s$^{-1}$)}\\
    & \multicolumn{2}{c}{Stack} & Simulated &  \multicolumn{2}{c}{Stack} & Simulated&  \multicolumn{2}{c}{Stack} & Simulated\\
    & Value & {Std Dev} & & Value &  Std Dev & & Value &  Std Dev &\\ 
  \hline

1a  Bright & 93.98&0.22 & 100.0 & 429.1&0.91 & 400.0 & 42.38&0.07 & 42.57\\
1a Faint & 0.477 &0.026 & 0.5 & 410.5&20.82 & 400.0 & 0.206&0.005 & 0.212\\
[0.2cm]
1b Bright & 99.92&0.025 & 100.0 & 403.8&0.25 & 400.0 & 42.58&0.021 & 42.57\\
1b Faint & 0.496&0.019 & 0.5 & 408.3&20.93 & 400.0 & 0.213&0.010  & 0.212\\
[0.2cm]
2a Bright & 93.38&4.630 & 99.33 & 429.2&0.971 & 400.0 & 42.12&2.081 & 42.29\\
2a Faint & 0.478&0.046 & 0.498 & 411.0&22.95 & 400.0 & 0.205&0.017 & 0.212\\
[0.2cm]
2b Bright & 93.92&5.71 & 99.85 & 429.5&4.59 & 400.0 & 42.39&2.60 & 42.51\\
2b Faint & 0.472&0.042 & 0.500 & 429.1&32.35 & 400.0 & 0.211&0.018 & 0.213\\
[0.2cm]
3 Bright & 90.20&1.59 & 100.0 & 637.0&47.87 & 592.1 & 60.33&5.18 & 63.03\\
3 Faint & 0.462&0.020 & 0.5 & 599.9&51.05 & 601.6 & 0.293&0.023 & 0.320\\
[0.2cm]
4 Bright & 90.49&5.34 & 100.3 & 645.1&41.73 & 596.4 & 61.31&5.85 & 63.71\\
4 Faint & 0.452&0.038 & 0.500 & 636.6&68.10 & 593.8 & 0.301&0.033 & 0.316\\
[0.2cm]
5a Bright & 92.84&0.308 & 100.0 & 430.4&0.861 & 400.0 & 42.52&0.082 & 42.57\\
5a Faint & 0.464&0.026 & 0.5 & 429.6&25.16 & 400.0 & 0.210&0.014 & 0.212\\
[0.2cm]
5b Bright & 92.43&0.266 & 100.0 & 430.6&1.156 & 400.0 & 42.33&0.065 & 42.57\\
5b Faint & 0.467&0.024 & 0.5 & 433.7&30.83 & 400.0 & 0.213&0.012 & 0.212\\
[0.2cm]
5c Bright & 92.04&0.273 & 100.0 & 430.4&0.932 & 400.0 & 42.14&0.098 & 42.57\\
5c Faint & 0.468&0.025 & 0.5 & 421.0&29.11 & 400.0 & 0.206&0.016 & 0.212\\
[0.2cm]
6a Bright & 95.29 & 2.88 & 100.0 & 428.6 & 1.003 & 400.0 & 41.24 & 1.25 & 42.57 \\
6a Faint & 0.451 & 0.051 & 0.5 & 362.3 & 30.57 & 400.0 & 0.169 & 0.014 & 0.212 \\
[0.2cm]
6b Bright & 91.11 & 2.72 & 100.0 & 428.6 & 1.109 & 400.0 & 39.42 & 1.16 & 42.57\\
6b Faint & 0.408 & 0.052 & 0.5 & 334.7 & 35.69 & 400.0 & 0.140 & 0.013 & 0.212 \\
[0.2cm]
7 Bright & 95.40&4.776 & 99.44 & 415.6&1.698 & 400.0 & 42.13&2.087 & 42.34\\
7 Faint & 0.496&0.052 & 0.507 & 410.7&32.31 & 400.0 & 0.214&0.025 & 0.216\\
[0.2cm]
8a Bright & 90.88&4.353 & 99.81 & 633.5&46.20 & 596.3 & 63.05&5.564 & 63.35\\
8a Faint & 0.438&0.087 & 0.507 & 629.7&87.33 & 611.0 & 0.289&0.040 & 0.330\\
[0.2cm]
8b Bright & 90.77&4.788 & 100.8 & 639.0&48.93 & 599.8 & 63.58&6.389 & 64.39\\
8b Faint & 0.411&0.061 & 0.497 & 651.6&111.1 & 608.0 & 0.286&0.044 & 0.322\\
[0.2cm]
9 Bright & 94.00&0.21 & 100.0 & 429.3&0.92 & 400.0 & 42.86&0.07 & 42.57\\
9 Faint & 0.475&0.021 & 0.5 & 410.5&17.98 & 400.0 & 0.205&0.006 & 0.212\\
[0.2cm]
\hline

\end{tabular}
\label{table:result}

\begin{flushleft}
$(a)$ Data set number ID ; $(b)$ Average line amplitude from all the 100 simulations of the studied set, and from their corresponding stacks ; $(c)$ Average FWHM of the emission lines, from all the 100 simulations of the studied set, and from their corresponding stacks ; $(d)$ Average integrated flux from all the 100 simulations of the studied set, and from their corresponding stacks.
\end{flushleft}
}

\end{table*}

\subsection{Stacking results from 1D simulations} \label{res-1D}

In the coming sections we will be analyzing results from stacking the spectra data sets presented in section \ref{simu1D}. We studied the reconstruction of the amplitude of the line, the linewidth and the integrated flux for all 5 data sets as well as some additional parameters specific to each set.

\subsubsection{Diagnostic of redshift uncertainties - set10} \label{res-9}

\begin{figure}
\includegraphics[height=60mm]{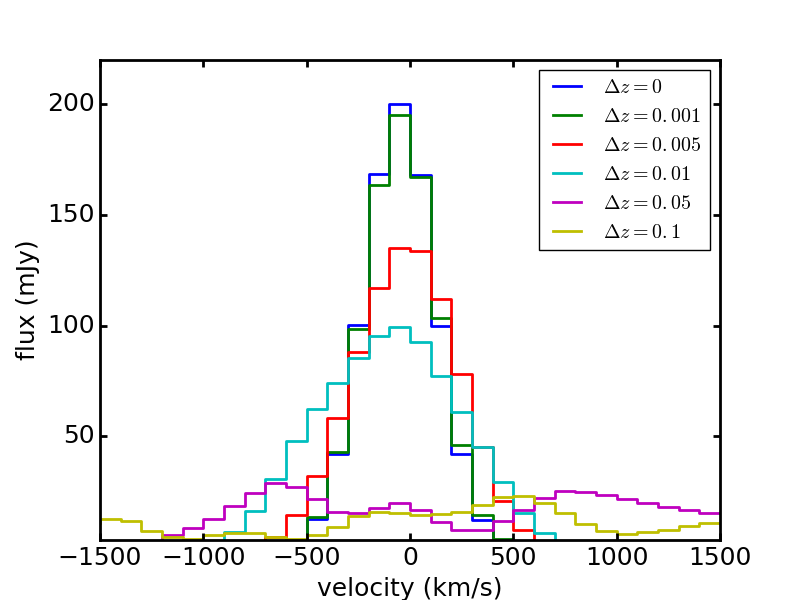}
\centering
\caption{Stack spectrum of 30 sources of width 400\,km\,s$^{-1}$, for different redshift uncertainties. $\Delta z=0.01$ already shows a $\sim50\%$ reconstruction, and rapidly dropping. We chose here a high SNR configuration (SNR before stacking $\sim 200$), to showcase the pure effect effect of redshift uncertainties on the stack. The corresponding velocity shifts are: $\sim 0, 36, 180, 360, 1800$ and $3600$\,km\,s$^{-1}$.}
\label{fig:dzTest}
\end{figure}

\begin{figure}
\includegraphics[height=60mm]{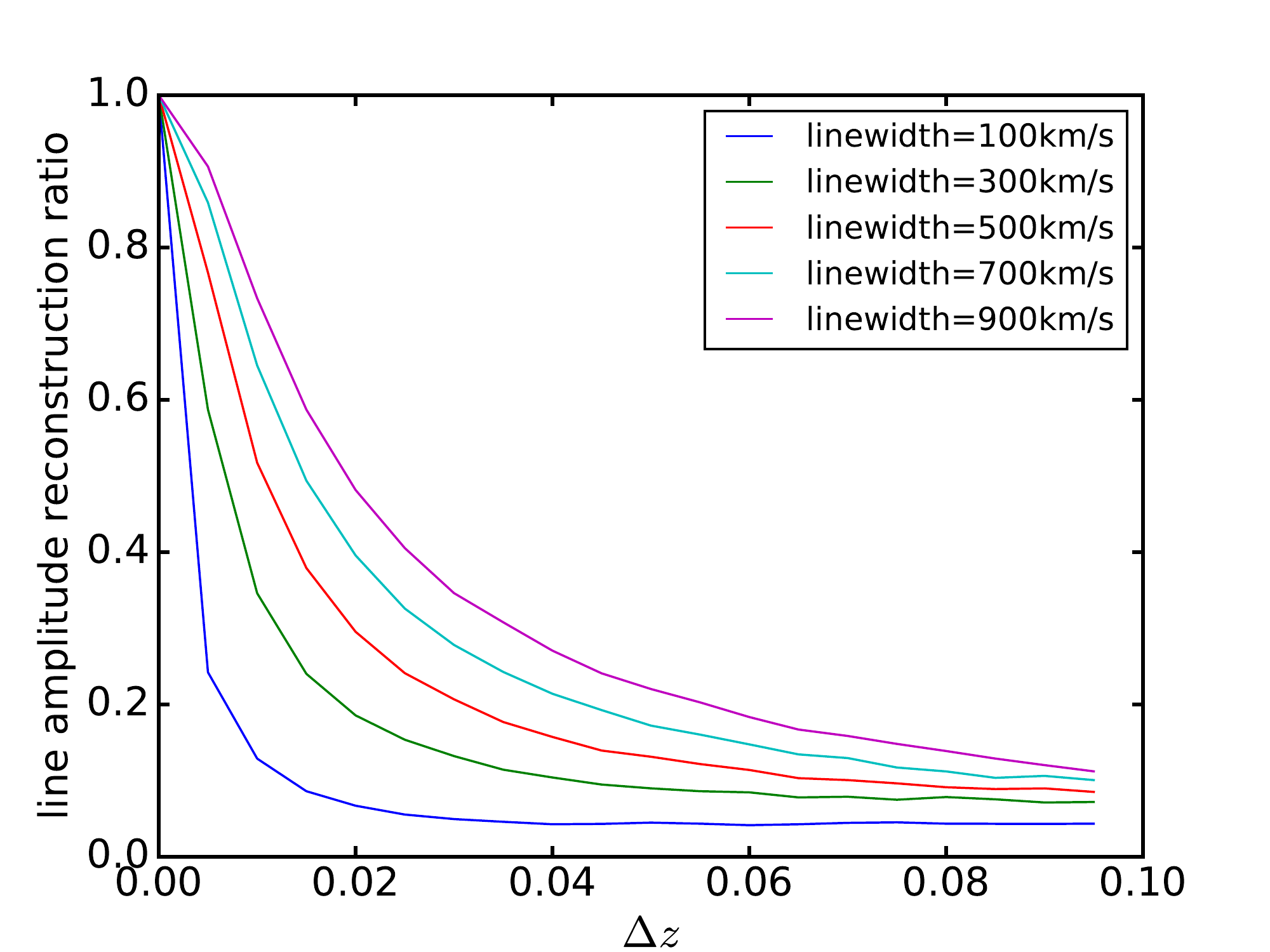}
\centering
\caption{Amplitude reconstruction ratio, for a stack of 30 noise-free sources, as a function of redshift uncertainties for different linewidth (all stacked sources are simulated with the same linewidth). Stacking very narrow lines (i.e. $\sim100\,$km\,s$^{-1}$) requires a very precise redshift. Results are averaged from 1000 realizations.}
\label{fig:dzflux}
\end{figure}

In this set we simulated lines with offset redshifts in order to study and estimate the effect of redshift uncertainties on the stacked line. It is important to note that the results of such a study depend on the average linewidth. Figure \ref{fig:dzTest} shows the average reconstructed line for different $\Delta z$ at a given linewidth of 400\,km\,s$^{-1}$, it shows that as soon as the redshift uncertainty becomes larger than 0.01 the stacked line cannot be recovered. Showcasing the importance of redshift accuracy. Figure \ref{fig:dzflux} shows the clear relation between linewidth and goodness of the reconstruction, when confronted to redshift uncertainties: if stacking high velocity lines, the effect of redshift uncertainty on the reconstruction will be lowered. { Alternatively reducing the spectral resolution would also mitigate the effect of redshift uncertainty on the reconstructed flux, at the expense of accuracy on the line profile measurement.} Results values are summarized in Table \ref{table:deltaZ1}.

\subsubsection{Double peaked spectrum - set11a} \label{res-10a}

This set studies a different spectral signature, of a double peak spectrum. The double peaked line is described by two Gaussians separated by a velocity difference, $D$, with 200\,km\,s$^{-1}<\,D\,<$\,600\,km\,s$^{-1}$. Both lines have the same width which is drawn at random between 100\,km\,s$^{-1}$ and 700\,km\,s$^{-1}$. Each stack consists of 30 spectra and is repeated for 1000 realizations. As shown in Table \ref{table:doublePeak-result} stacking allows for a good reconstruction of all three parameters: 91\%, 97\% and 94\% for the line amplitude, distance between the peaks and linewidth in the high SNR case and low SNR cases. The integrated flux is also well reconstructed at $\sim$ 95\%. The standard deviation of the studied parameters is low ($\sim 5\%$) hence one can expect a good degree of confidence when studying similar cases with no redshift uncertainties. 

\subsubsection{Double peaked spectrum with a $\Delta z$ - set11b} \label{res-10b}

\begin{figure}
\includegraphics[height=60mm]{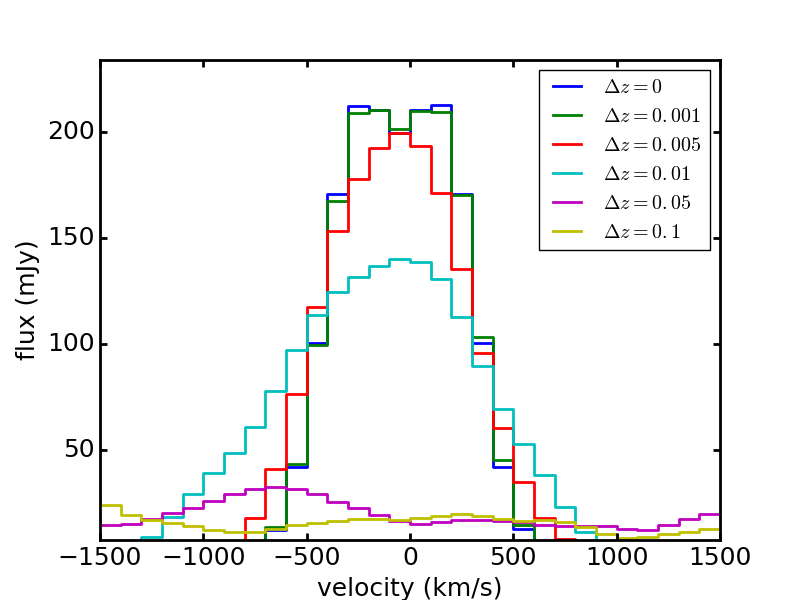}
\caption{Stack spectrum of 30 sources with double peak profile for different redshift uncertainties (both peaks are Gaussian, with a width of 400\,km\,s$^{-1}$, and a distance of 400\,km\,s$^{-1}$ between the two peaks). An uncertainty higher than $0.001$ already dilutes the two peaks, showing that similar profiles will be extremely hard to reconstruct through stacking. {Similarly to Figure \ref{fig:dzTest} we chose a high SNR ($\sim 200$ before stacking) to showcase the pure effect of redshift uncertainties. Corresponding velocity shifts are: $\sim 0, 36, 180, 360, 1800$ and $3600$\,km\,s$^{-1}$.}}
\label{fig:doublePeakDZ}
\end{figure}

Data set 11b is similar to set 11a but focuses on the effect of redshift uncertainties on the stack. Linewidth and distance between the peaks are kept constant, both at 400\,km\,s$^{-1}$.  Results presented are the average and standard deviation of 1000 realizations. Figure \ref{fig:doublePeakDZ} shows that, as soon as redshift uncertainty becomes worse than 0.001 the double peak feature is no longer distinguishable. This implies that such spectral signatures will be very hard to observe using stacking, and will require a very good redshift accuracy. Furthermore, the reconstruction of such a feature will also depend on the distance between the peaks. If the separation between the peaks is higher they will be easier to discern. Table \ref{table:doublePeak-result} shows the best fit of the stack, using a double peak fit. { While the reconstruction of all three parameters is near perfect at $\Delta z=0$, from $\Delta z\sim0.05$ the line cannot be recovered (see Figure \ref{fig:doublePeakDZ})}. One should note that the integrated flux is not impacted as much by the redshift uncertainties as the other parameters, thus in such cases it is advised to focus on the integrated flux rather than other line parameters. 

\begin{table*}
\centering


{
\caption{Stacking results from 1D simulated data set9. Presented results are obtained by averaging 1000 stacks. Parameters are obtained through fitting. Presented errors are computed standard deviation of the given parameter in the 1000 stacks. }
\begin{tabular}{ccccccccc}
  \hline
  data set&$\Delta z^{(a)}$&\multicolumn{2}{c}{Mean line amplitude$^{(b)}$ }&\multicolumn{2}{c}{Mean linewidth$^{(c)}$} &\multicolumn{2}{c}{Mean integrated flux$^{(d)}$}\\
  & &\multicolumn{2}{c}{(mJy)} &\multicolumn{2}{c}{(km\,s$^{-1}$)} &\multicolumn{2}{c}{(mJy\,km\,s$^{-1}$)}\\
   & & Value &  Std Dev & Value &  Std Dev & Value & Std Dev  \\
   
  \hline

10 Bright&0&200.01&0.13&400&2.1&83289&53\\ 
10 Faint&0&1.02&0.13&400&64&416&52\\ 
[0.2cm]
10 Bright&0.001&195.66&1.11&408&2.3&83027&151\\ 
10 Faint&0.001&1.01&0.13&407&67&416&51&\\ 
[0.2cm]
10 Bright&0.005&138.91&9.71&579&44&75592&2219\\ 
10 Faint&0.005&0.72&0.12&581&123&381&50\\ 
[0.2cm]
10 Bright&0.01&87.73&10.63&924&127&57866&4905\\ 
10 Faint&0.01&0.47&0.14&910&289&292&57\\ 
[0.2cm]
10 Bright&0.05&-&-&-&-&-&-\\ 
10 Faint&0.05&-&-&-&-&-&-\\ 
[0.2cm]
10 Bright&0.1&-&-&-&-&-&-\\ 
10 Faint&0.1&-&-&-&-&-&-\\ 
[0.2cm]
13 Bright&0&200.0&0.11&399.99&0.30&85155&158\\ 
13 Faint&0&0.99&0.42&398&74&421&168\\ 
[0.2cm]
\hline
\end{tabular}
\label{table:deltaZ1}

\begin{flushleft}
$(a)$ Average redshift uncertainty, leading to uncertainty of the line-center position; $(b)$ Average resulting stacked line amplitude; $(c)$ Average resulting stacked line FWHM; $(d)$ Average resulting stacked line integrated flux.
\end{flushleft}
}
\end{table*}

\begin{table*}
\centering

{
\caption{Stacking results from 1D simulated data sets with double peak profiles (set11a and set11b). Presented results are obtained by averaging 1000 stacks. Parameters are obtained through fitting. Presented errors are computed standard deviation of the given parameter in the 1000 stacks.}
\resizebox{\textwidth}{!}{\begin{tabular}{cccccccccccccccc}
  \hline
  Data Set$^{(a)}$ &$\Delta z^{(b)}$&
  \multicolumn{3}{c}{Mean line amplitude$^{(c)}$} &
  \multicolumn{3}{c}{ Mean peak distance$^{(d)}$} &
  \multicolumn{3}{c}{ Mean linewidth$^{(e)}$} &
  \multicolumn{4}{c}{ Mean integrated Flux$^{(f)}$}\\
  
  & &
  \multicolumn{3}{c}{(mJy)} &
  \multicolumn{3}{c}{ (km\,s$^{-1}$)} &
  \multicolumn{3}{c}{ (km\,s$^{-1}$)}&
    \multicolumn{4}{c}{ (mJy\,km\,s$^{-1}$)}\\

   & & \multicolumn{2}{c}{Stack} & Simulated & \multicolumn{2}{c}{Stack} & Simulated& \multicolumn{2}{c}{Stack} & Simulated&
   \multicolumn{2}{c}{Stack} & \multicolumn{2}{c}{Simulated}\\
   & & Value & Std Dev &  & Value & Std Dev & & Value & Std Dev & & Value & Std Dev & Value & Std Dev\\
   
  \hline

11a&0&182.4&7.2&200.0&413.7&23.9&400.0&428.2&36.7&400.9&168778 & 9490 & 172595 & 8841 \\ 
11a&0&0.91&0.21&1.0&414.3&334.7&400.0&427.4&157.4&399.9&768 & 263 & 805 & 47\\ 
[0.2cm]
11b&0&199.9&0.13&200.0&400.0&0.15&400.0&400.0&0.28&400.0&169689 & 64& 170301&- \\
11b&0&1.007&0.170&1.0&400.0&419.7&400.0&397.8&97.62&400.0&850 & 65 & 851&-\\  
[0.2cm]
11b&0.001&195.6&2.33&200.0&400.0&0.16&400.0&408.9&4.77&400.0&169566& 71 & 170301&-\\
11b&0.001&0.98&0.17&1.0&400.7&337.3&400.0&409.7&99.19&400.0&849 & 66 & 851&-\\ 
[0.2cm]
11b&0.005&136.9&29.21&200.0&399.9&3.99&400.0&584.2&90.00&400.0&164482 & 1604& 170301&-\\
11b&0.005&0.68&0.216&1.0&399.6&336.0&400.0&586.4&154.8&400.0&822 & 66 & 851&-\\
[0.2cm]
11b&0.01&84.72&46.83&200.0&400.2&114.4&400.0&944.4&221.8&400.0&145492& 5947& 170301&-\\
11b&0.01&0.41&0.26&1.0&337.3&336.5&400.0&985.0&247.7&400.0&726 & 73 & 851&-\\
[0.2cm]
11b&0.05&-&-&200.0&-&-&400.0&-&-&400.0&-&-&170301&-\\ 
11b&0.05&-&-&1.0&-&-&400.0&-&-&400.0&241&92&851&-\\ 
[0.2cm]
11b&0.1&-&-&200.0&-&-&400.0&-&-&400.0&-&-&170301&-\\
11b&0.1&-&-&1.0&-&-&400.0&-&-&400.0&124&88&851&-\\ 

\hline
\end{tabular}
\label{table:doublePeak-result}
}
\begin{flushleft}
 $(a)$ Data set number ID; $(b)$ Average redshift uncertainty, leading to uncertainty of the line-center position; $(c)$ Average line amplitude, from the stack and from simulations; $(d)$ Average distance between the two peaks, from the stack and from simulations; $(e)$ Average single line FWHM (both lines have the same width), from the stack and from simulations; $(f)$ Average line integrated flux, from the stack and from simulations.
 \end{flushleft}
 }
\end{table*}

\subsubsection{Outflows - set12} \label{res-outflow}

Data set12 was built to study cases with spectral signature of outflows. Spectra consist of two components, one main line with an amplitude of 2000\,mJy and a width of 400\,km\,s$^{-1}$, and an outflow component with an amplitude of 200\,mJy and a width of 1000\,km\,s$^{-1}$ in the high SNR case. Amplitudes are set to 10 mJy and 1 mJy for the main and outflow component respectively in the low SNR case (linewidth are the same as in the high SNR case). Results are obtained through fitting the stacks with two Gaussian components (results are averaged from the 1000 realizations). {In the low SNR sets the amplitude of the main component is fixed when fitting. This is done to avoid cases where good fitting is achieved with a brighter broad component and a fainter main line, leading to a much higher uncertainty on the broad component amplitude reconstruction (of order $\sim$\,80\%).} 

Once again \textsc{LineStacker} allows for a good reconstruction, with about $\sim99\%$ retrieval of the parameters for both components in both SNR configuration {(assuming the fitting method described above in the low SNR configuration)}. It should be noted however, that, when stacking lines of different linewidths, the variation in linewidths needs to be accounted for, the method would otherwise be biased in finding outflows. To do so a spectral rebinning method can be applied to the data pre-stacking (see section \ref{rebin} and \citealt{Stanley2019}).

\begin{table*}
\centering
{
\caption{Stacking results from 1D simulated containing an outflow component (data set12). Presented results are obtained by averaging 1000 stacks. The characteristics of both components are obtained through Gaussian fitting with two Gaussian. Presented errors are computed standard deviation of the given parameter in the 1000 stacks}
\begin{tabular}{cccccccccccc}
  \hline

\multicolumn{3}{c}{Mean main line amplitude$^{(a)}$} &
  \multicolumn{3}{c}{ Mean outflow amplitude$^{(b)}$} &
  \multicolumn{3}{c}{ Mean main line width$^{(c)}$} &
  \multicolumn{3}{c}{ Mean outflow width$^{(d)}$}\\
  \multicolumn{3}{c}{(mJy)} &
  \multicolumn{3}{c}{(mJy)} &
  \multicolumn{3}{c}{ (km\,s$^{-1}$)} &
  \multicolumn{3}{c}{ (km\,s$^{-1}$)}\\
  \multicolumn{2}{c}{Stack} & Simulated & \multicolumn{2}{c}{Stack} & Simulated & \multicolumn{2}{c}{Stack} & Simulated& \multicolumn{2}{c}{Stack} & Simulated \\
  Value & Std Dev & & Value & Std Dev &  & Value & Std Dev & & Value & Std Dev & \\
  \hline\\
1999.&0.46&2000.0&200.0&0.47&200&399.9&0.07&400.0&999.9&1.086&1000.0  \\ 
10.00& - &10.0&0.99&0.13&1.0&401.5&10.9&400.0&984.5&131.1&1000.0 \\

	\hline

	\end{tabular} \\
\label{table:outflow-result}

\begin{flushleft}
$(a)$ Average main component amplitude, from the stack and from simulations ; $(b)$ Average second component amplitude, from the stack and from simulations ; $(c)$ Average main component line FWHM, from the stack and from simulations ; $(d)$ Average second component line FWHM, from the stack and from simulations.
\end{flushleft}	
}
\end{table*}

\subsubsection{Lines on the edge - set13} \label{res-LinesOnEdge}

\begin{figure}
\centering
    \includegraphics[width=\linewidth]{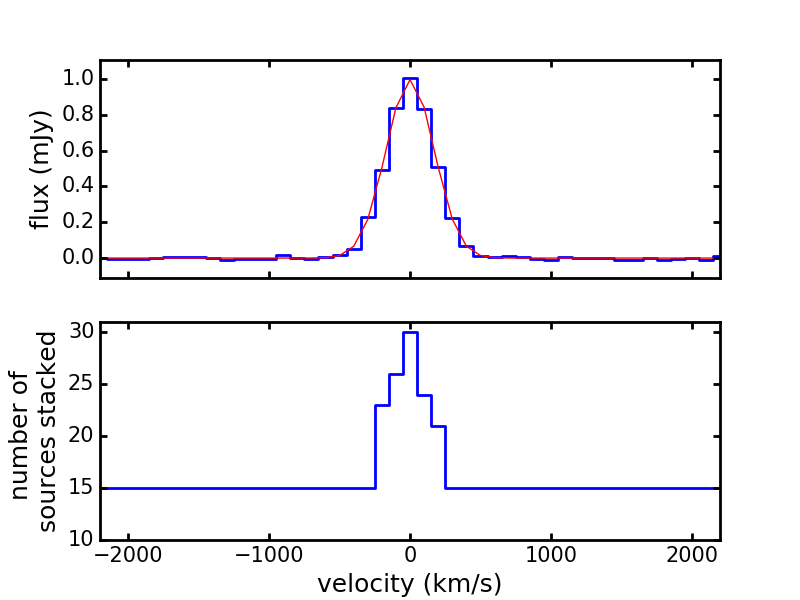}
\caption{Average results from 1000 realizations of stacking 30 sources from set13 (in the low SNR configuration). {\it (top)} Resulting average stack and the corresponding Gaussian fit (in red). {\it (bottom)} Number of sources stacked at each velocity bin. Due to the nature of the data stacked the number of sources quickly drops to half of the sources when moving away from the central channels.} 
\label{fig:LineOnEdge}
\end{figure}

Data set13 was built to diagnose cases where all lines would lie on the edge of the spectral window. Lines have the same properties as data set10 when $\Delta z=0$. Figure \ref{fig:LineOnEdge} shows an example from such a stack product. One can see that the numbers of sources stacked rapidly drops on both sides of the central channel. However reconstruction rates stay close to perfect for both noise configurations for all parameters, see Table \ref{table:deltaZ1}. {It should nonetheless be noted that, while average results are consistent with data set10 at $\Delta z=0$, the standard deviation is significantly higher than set10 in the low SNR regime (of order $\sim 3$ times higher for both the amplitude and the integrated flux), indicating a larger spread in the expected results.}

\section{Discussion} \label{discussion}

\subsection{Choice of parameters.} \label{disc:simParam}

In the following subsections we justify the choice of parameters for our simulated data-sets. 

\subsubsection{Array configuration}

Even if a specific array configuration has been chosen for most simulations (ALMA cycle 6 configuration 2), the method has been extensively tested in other array configurations (not presented here), as well as with other choice of interferometer -- as showed through data sets 8a and 8b, simulating VLA observations -- and neither should have any impact on the performance of the algorithm.

\subsubsection{Number of sources} \label{disc:sourcesNumber}

For every simulation, we chose to stack 30 sources. The number 30 has been chosen as an intermediate number, allowing for both a good noise reduction ($\sqrt{30} \sim 5.5$) and a satisfying computing time. This number is typical for small, but statistically significant, samples. A lower number would imply too important statistical fluctuations and, on the other hand, a higher number of sources \citep[samples as big as a few thousands e.g. ][can be expected]{Dole2006} would not change the conclusion from our analysis, simply allowing identification of fainter signal. 

\begin{figure}
\includegraphics[height=60mm]{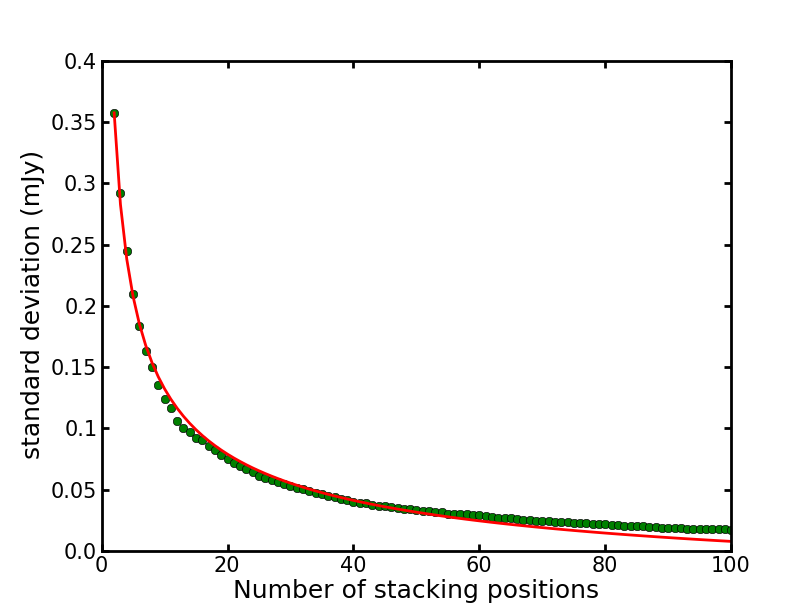}
\centering
\caption{standard deviation in the stack of empty cubes (simulated through SIMALMA) as a function of number of cubes stacked. Red line indicates the theoretical $1/\sqrt{N}$ drop off. Presented results are averaged from 100 realizations.}
\label{fig:sqrtN}
\end{figure}

To verify the good behavior of noise reduction as a function of number of sources stacked, we stacked an increasing number of {empty fields (i.e. without sources, containing just noise)}. The fields have similar noise properties as in the rest of our simulations.  Computing for each new number of stack positions the standard deviation across the map, see Figure \ref{fig:sqrtN}. The noise reduction is in good accordance with theoretical predictions (showing an 80\% noise reduction for 30 sources, similar to the 82\% theoretically expected)  confirming the relevance of the number of sources in our stacks.

\subsubsection{Number of simulations} \label{disc:simNumber}

Each data set has been simulated, stacked, and analyzed 100 times in the case of 3D data sets, and 1000 in the case of 1D data sets. The number of realizations has been chosen for practical reasons, limiting computation time, it should however be noted that the total number of resulting sources is high enough to provide statistically relevant numbers. This can be seen from the uncertainties on the reconstructed parameters -- coming directly from the spread of results in the stacking iterations -- that are small compared the found values.

\subsubsection{High and low SNR sets} \label{disc:SNRSets}

Every 3D simulated data set was run both in a high and a low SNR configuration. 
While high SNR data sets reveal the ideal behavior of our algorithm and show that it performs well, stacking will be mostly used on low SNR data. It was therefore essential to show that our tool was successful in such cases as well. Flux values have been chosen arbitrarily, mimicking typical values. The amplitude of the line in low SNR sets has been chosen such that individual lines would be of order of the noise, thus impossible to detect individually but strong enough to be studied once 30 of such sources are stacked.

\subsubsection{Source size} \label{disc:size}

Most simulated data-sets have been performed simulating point sources. It is not uncommon that high-redshift galaxies are seen as point sources because of the high resolution needed to resolve such objects, and working with point sources allows to have a more controlled data-set. We have however studied extended sources as well, in data-sets 6a and 6b, showing that no substantial effect would come from stacking extended sources instead of point sources.
However, only homogeneous cases have been tested (where sources were either all point like or all extended). Furthermore, when studying extended sources all sources had similar sizes. It may happen that stacked sources have very different sizes, resulting in an inhomogeneous spatial distribution in the stack product. In this case one should be prudent when extracting quantitative conclusions from such an analysis (such as sizes, density, spatial distribution etc.). Using statistical methods like bootstrapping or subsampling should prove useful to diagnose such cases. Furthermore, if the sources extent is identifiable pre-stacking, extracting and stacking directly the spectra -- using the 1D stacking module of \textsc{LineStacker} -- could allow to get past these biases \citep[see][for an example use of combination of 1D and 3D stacking]{Stanley2019}.

\subsection{Impact of redshift uncertainty on stack} \label{disc:dz}

When stacking spectral line data, the two most important properties to know are the position and redshift of the source. Uncertainties in the astrometry will result in the stacked source appearing more extended than in reality \citep[see e.g.][for further discussion on astrometric uncertainties]{Lindroos2016,Lindroos2018}. 
In sections \ref{res-9} and \ref{res-10b}, we have shown that uncertainties on the redshift have a significant impact on the stacked line result as well as on the potential reconstruction of the average line profile. The {three} most common cases of high-redshift galaxies, where the uncertainties could be too large are (i) when using photometric redshifts for a large number of galaxies, (ii) if relying on the ultra-violet, broad emission lines of quasars, as these can often be shifted relative to the  velocity of the quasar host \citep[e.g.][]{Coatman2017}, {and (iii) when studying Lyman-alpha emitters because their peak emission is known to be offset from the systemic redshift \citep[e.g.][]{Shapley2003, Rakic2011}} . 

An uncertainty of $\Delta z=0.05$ results in the amplitude of a line of 400 km\,s$^{-1}$, with a spectral accuracy of 100 km\,s$^{-1}$, to be recovered at only $\sim$ 20\%. However wider lines will show a better reconstruction (see Figure \ref{fig:dzflux}) at similar redshift uncertainties.
\cite{Bellagamba2012} have shown that one can typically expect $\langle \frac{\sigma_{\rm z}}{1+z}\rangle<0.05$ from photometric redshift of weak lensing surveys, which at $z\sim6$ corresponds to $\Delta z < 0.35$. Combined with our analysis this shows that using photometric redshift to stack lines from high-redshift objects should not be viable. 

Other examples of significant offsets can come from quasars and powerful AGN, where the redshift is determined by lines that are several thousands 
km\,s$^{-1}$ wide and can be offset due to outflows. This can result in shifts that are so large that the line might be shifted out of the interferometric bands. An example of this is W0410-0913 shown in \citet{Fan2018} where even the optical redshift was a few thousands km\,s$^{-1}$ offset. 

\subsection{Gaussian fitting to recover stacked line parameters} \label{disc:underEval}

When stacking lines with a wide range of line widths the resulting stacked line ends up not being Gaussian. Therefore, using Gaussian fitting to recover the line parameters will yield biased results. This effect is especially relevant if one is interested in retrieving the line shape. It should be noted however that this effect can be mitigated depending on the spread in line widths. For example, in simple cases such as the one given in Figure \ref{fig:badFit}, fitting a single Gaussian line profile is a good approximation for the used data sets. However, in more complicated or realistic cases, as for example the case given in  Appendix \ref{ex:CubeExtended} and Figure \ref{fig:noRebinFit}, one can see that deriving parameters using single Gaussian fitting would yield extremely biased results. In addition, it is important to keep in mind that, even if not interested in line shape, using Gaussian fitting to retrieve the stacked line amplitude will lead to its systematic under-estimation (see Figure \ref{fig:badFit}).

\subsection{Stacking in the {\it uv}-plane} \label{disc:uv}

In this paper we focus on stacking in image-plane. \citet{Lindroos2015}, conducted a systematic comparison between stacking interferometric data in the image-plane and in the {\it uv}-plane. The two approaches yield similar results, but the {\it uv}-stacked results are more robust for a range of cases; in some cases the difference was a few percent, while in other cases the improvement was up to 15\%. 
The cases where {\it uv}-stacking has the potential to make a difference are: 
(i) when the {\it uv}-coverage differs between data sets: this can affect the stacked result. For example, if stacking in the image-plane the data should be imaged to the same resolution, however, this might not always be easy, depending on the array configuration. Furthermore, if the stacked sources are faint, i.e. below the few sigma r.m.s. of the data, the sources will not have been cleaned in the imaging and therefore the stacked image will include a stacked version of the dirty beams. The potential complications of these effects are avoided when stacking directly in the {\it uv}-plane as the imaging is done afterwards. (ii) When bright sources are present in the data, these have to be removed from the data before stacking, both for {\it uv}- and image-plane. This is normally done using the \textsc{clean} algorithm, and the residuals from this
can leave noise and cleaning artifacts in the data. Often this affects only the short baselines, and can hence be more easily identified and dealt with in {\it uv}-stacked data than in the image-plane stacked result.
(iii) When stacking extended source, it is possible to analyse the stacked {\it uv}-data for possible cleaning artifacts or problematic baselines that could affect size measurements, and it is possible to estimate the extend of the emission on the visibilities. 

In comparison between stacking in the {\it uv}-plane and image-plane, the 
image-plane stacking is computationally faster, in particular if also 
including statistical tests using Monte Carlo/resampling methods. { In addition image stacking allows for easier masking and is generally more intuitive to work with. It has been shown in \citet{Lindroos2015} to yield satisfying results when the imaging can be done reliably. Finally, future interferometers such as the SKA, will likely not archive all visibility data after processing and imaging because of the enormous amounts of data produced.}

In this paper, we have focused on studying the stacking of spectral line
data, and this has a number of complications that are independent of whether the stacking is done in the image- or the {\it uv}-plane. Spectral line stacking can be carried out in both image- and {\it uv}-plane, and the differences in performance is expected to be the same as was found for continuum stacking in \citet{Lindroos2015}. 

{Finally, we note that if the line width of the stacked line is known a priori, an easy approach allowing a coarse version of line stacking in the {\it uv}-plane could be to treat the line channels as a continuum channel \citep[see e.g. ][Mendez et al. 2020, in press]{Fujimoto2019}. Through this approach, it is the line intensity that is stacked, while all potential information about line profile is not included.}

\subsection{\textsc{LineStacker} as a spectral analysis tool} \label{disc:usage}

While we used \textsc{LineStacker} solely to study emission lines from high-redshift galaxies, the tool is flexible and would perform equally well on other type of data in the range GHz to THz. It is probable that the tool performs equally well on lower frequencies, but has not been tested. Additionally, while all the tests focused on stacking interferometric data, it is possible to use it to stack non-interferometric data (this could possibly include optical integral field spectroscopic data, though we note that performance has not be tested). The tool could for example be used on stars, clusters, gas clouds, local galaxies, galaxy clusters or any other object. In principle any line data can be studied, and it would also be possible to stack different lines from the same object together (from different transitions for example), { provided that the sub-cubes spectral size is chosen as compact enough to avoid overlap between the different lines}. Besides, stacking absorption lines is theoretically identical and should yield similar results. It would also be possible in principle,  using Monte Carlo methods, to find better individual redshifts of the studied objects, by trying to optimize stacked line reconstruction. Indeed, stacked line amplitude should be maximized when all the target sources are stacked in phase, i.e. when all lines are stacked perfectly all in the velocity center. One could then look for the set of redshifts maximizing the stacked line amplitude, recovering, in fine, better individual redshifts while also optimizing line reconstruction.

\subsection{Spectroscopic stacking in the literature}

Spectroscopic stacking as a method has been used more and more in the past decade. When presented in the literature authors typically have not shown tests of their algorithm nor made it publicly available. Among other, stacking has been used to study H\textsc{i}, both in emission and absorption \citep{Murray2014,Murray2018}. H\textsc{i} line profiles are usually complex, and, while the redshift precision is usually good, such complexity will typically be diluted through stacking. Aside from H\textsc{i}, other molecules have been studied, such as CO \citep[e.g][in multiple transitions]{Decarli2016} or [C\,\textsc{ii}] \citep[e.g.][]{Decarli2018QSO,Bischetti2018,Stanley2019,Fujimoto2019}. 
In \citet{Decarli2016}, stacking was performed for a sample of high-$z$ galaxies, where the optical redshifts were uncertain by 200-1000\,km\,s$^{-1}$, yielding a low-significance detection. This highlights the challenge of stacking spectroscopic data without accurate redshifts or velocities. 
\citet{Bischetti2018} studied line profiles through stacking without rebinning their data. Nonetheless, they build physically selected sub samples (depending for example on the width of the lines) to reduce the effect of stacking sources with different linewidth. {In \citet{Stanley2019} we used \textsc{LineStacker} to search for outflows in high-redshift quasars. In addition to the main algorithm, we used some of tools included in \textsc{LineStacker} to improve our analysis: spectral rebinning to properly recover line profiles, as well as subsampling to identify the sources exhibiting the best outflow signature. Our analysis demonstrated the efficiency and flexibility of \textsc{LineStacker} to study faint emission at high redshift.}

\cite{Murray2014} used a different technique for edge treatment than the one presented in section \ref{edge}. In their algorithm they simply added zeros to fill spectral channels, when sources' lines were too close to the edge of the observation window. This has a bigger, unwanted, impact on the stacked result than the method used in \textsc{LineStacker}, because zeros added to the stack will, once averaged with the rest of the sources, artificially drive the edges of the stacked spectral window to values closer to zero. 

Using \textsc{LineStacker} would allow a fast, uniform and controlled way to do spectral stacking. It would also allow the use of statistical tools as well as data handling treatment in a more systematic and coherent way.


\section{Summary} \label{ccl}
We have carried out an extensive analysis of stacking of interferometric spectral line data using our new algorithm and tool \textsc{LineStacker}. We have used simulations of near-ideal and realistic cases of simulated data from two different interferometers, ALMA and VLA. All high SNR simulations emphasized the controlled behavior of our algorithm while low SNR simulations focused on noise reduction and proved the efficiency and usefulness of stacking.

We showed and justified the need for statistical tools, both pre-stacking (e.g. rebinning) and post-stacking (e.g. bootstrapping), to better understand stacking results as well as the distribution of parameters of the stacked population.

We find that knowing the redshift of the sources with a good precision is a necessary condition for a good line reconstruction. And that, for an average linewidth of 400\,km\,s$^{-1}$, a redshift uncertainty below 0.01 implies a line reconstruction around 60\%, dropping to roughly 10\% at $\Delta z=0.1$. Furthermore, it has been shown that more complex spectral signatures will tend to be smeared out by uncertainties on the redshift, and that hence, in most cases, more complex spectral signatures will disappear when stacked (either through averaging with other, different, spectral shapes, or because of the homogenization due to redshift error).

In addition, we showed that stacking Gaussian lines with different linewidth results in a non Gaussian shaped stack, leading to a possible misinterpretation of the fitted result. This can be fixed by rebinning the spectra, if individual linewidths are identifiable before stacking. Such spectral configuration are especially problematic when trying to identify line profiles.

{
With the significantly improved capabilities of modern radio and mm interferometers in combination with deep optical and near-infrared surveys, the use of stacking is becoming increasingly relevant. As seen in the literature, there is a growing interest to exploring the faint, often individually undetected sources through stacking. However the number of public tools available are still limited. \textsc{LineStacker} provides the community with increased opportunity for optimal synergy between modern telescopes and the large astronomical surveys. Enabling multi-wavelength studies is necessary also for faint sources in order to establish a complete understanding of the chosen population. }

\textsc{LineStacker} is open source and open access. It can be downloaded at \textit{https://www.oso.nordic-alma.se/software-tools.php}. The tool is provided with examples and documentation.

\section{Acknowledgements}
We thank the anonymous referee for their instructive comments for the improvement of this paper.
We thank John Conway for extensive discussion on the project. 
We thank the staff of the Nordic ALMA Regional Center node for their support. 
KK acknowledges support from the Swedish Research Council (2015-05580). 
JBJ thanks Martin Zwaan for helpful discussions. 

\bibliography{myBib}



\appendix
\section{Example use of \textsc{LineStacker}}

All the following examples are presented in more detail in the \textsc{LineStacker} library, including the lines of code, and the associated data sets, required to run them.

\subsection{Cube basic example} \label{ex:CubeBasic}

The first example shows a basic use \textsc{LineStacker}. In many cases stacking is done for sources where the target lines are not individually detected and where the redshift or systemic velocity has been determined through other means (e.g. in another wavelength range). For example, this could be measuring the stacked CO line of optically detected galaxies, where the redshift is determined from optical spectroscopy. 

\begin{figure}
\centering
    \includegraphics[width=\linewidth]{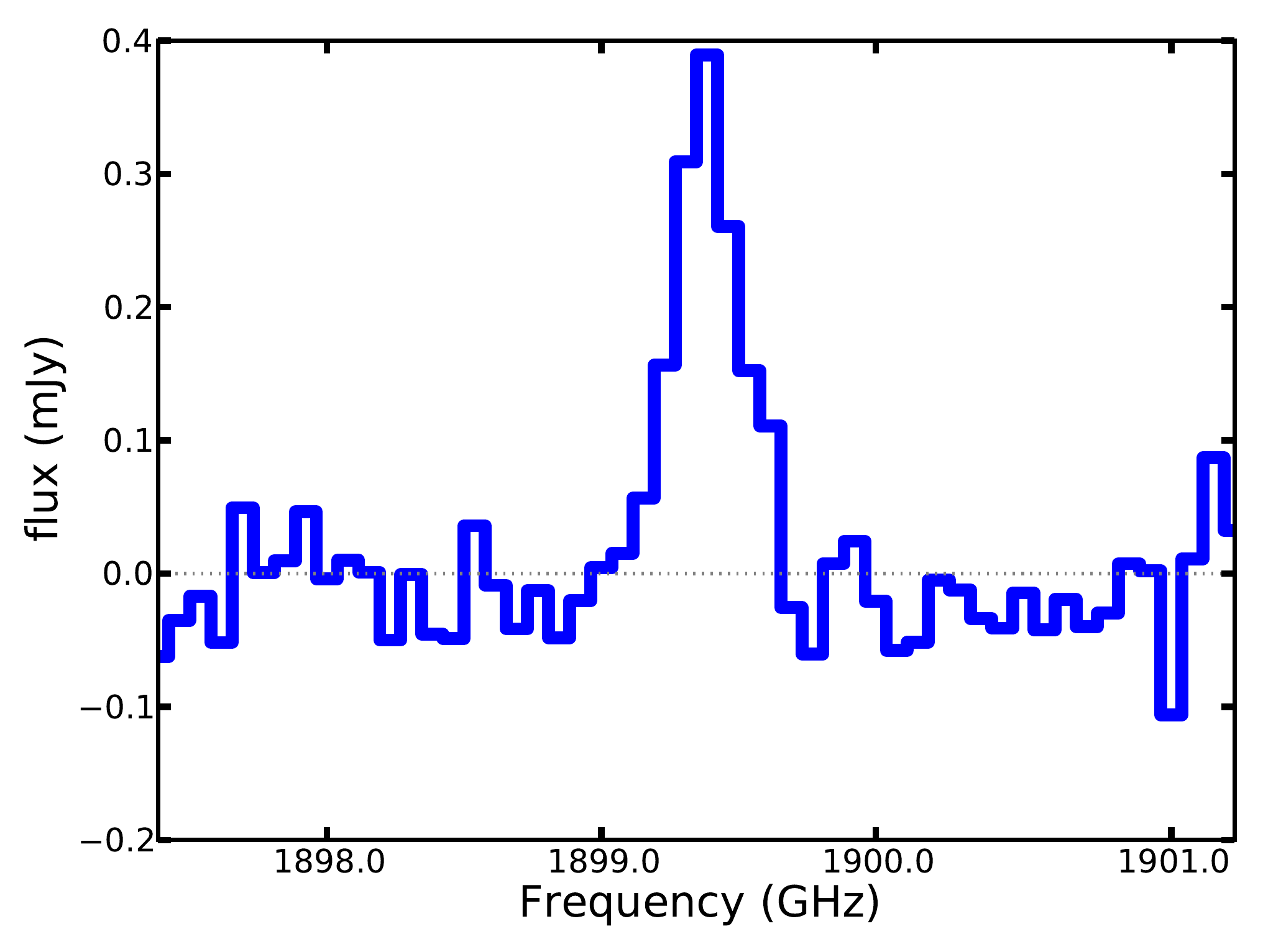}
    \includegraphics[width=\linewidth]{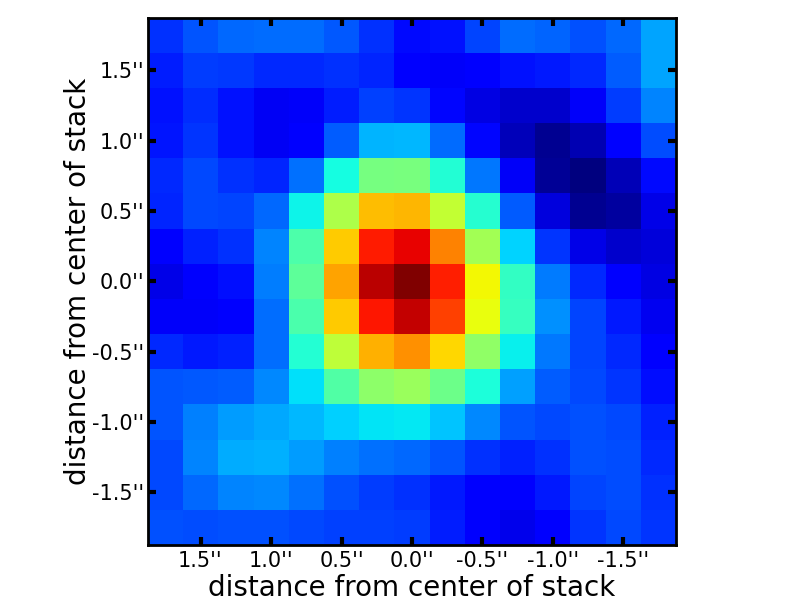}
\caption{Results from the example of a basic stacking. 
Top: Spectrum extracted from central pixel of the stack. Note that the frequency values are arbitrary. 
Bottom: Moment-0 map of the mean stack. 
\label{fig:basicCube}}
\end{figure}

The input required together with the data cubes are source coordinates and line center, which should be given in as a file. The coordinates are given in J2000 R.A. and Dec., while the line center is tabulated with the source redshift and rest-frame frequency of the line. We note that several methods for determination of the line center are available, for example central bin index and central velocity -- full description of the line center identification methods are available in the \textsc{LineStacker} documentation.  
Here we show an example with 30 cubes, each containing a point source in its center. The sources spectra consist solely of an emission line and some noise. The amplitude of the lines has, on average, the same amplitude as the noise (similarly to set1a in the low SNR regime).  Mean stacking is performed on the cubes, and the result can be seen on Figure \ref{fig:basicCube}.

\subsection{Cube extended example} \label{ex:CubeExtended}
Most high-$z$ galaxies will display varying line profiles and line widths depending on properties such as mass and orientation. Here follows an example expanded from the previous subsection, where we use simulated data of sources with random line widths. As discussed in the main body of the paper, stacking sources of varying line widths can affect the final result. The present example details the usage of 
the rebinning method coupled to stacking. We note that using rebinning requires some prior realistic knowledge of the line width, this could for example be stacking of fainter isotopologue lines under the assumption that the line width is similar to individually detected main isotopologue lines (e.g.\ $^{12}$CO compared to $^{13}$CO; e.g.\ Mendez et al., in prep.) or the search for  high-velocity outflows \citep[e.g.][]{Decarli2018QSO,Stanley2019}.

The data used here consists of 50 cubes, with one point source each in their center. The sources spectra consist of a bright (SNR $\sim100$) Gaussian line with random width. The randomization of the width is operated through Gaussian random centered around $200$\,km\,s$^{-1}$, with a minimum value of $50$\,km\,s$^{-1}$. 

\begin{figure}
\includegraphics[width=\linewidth]{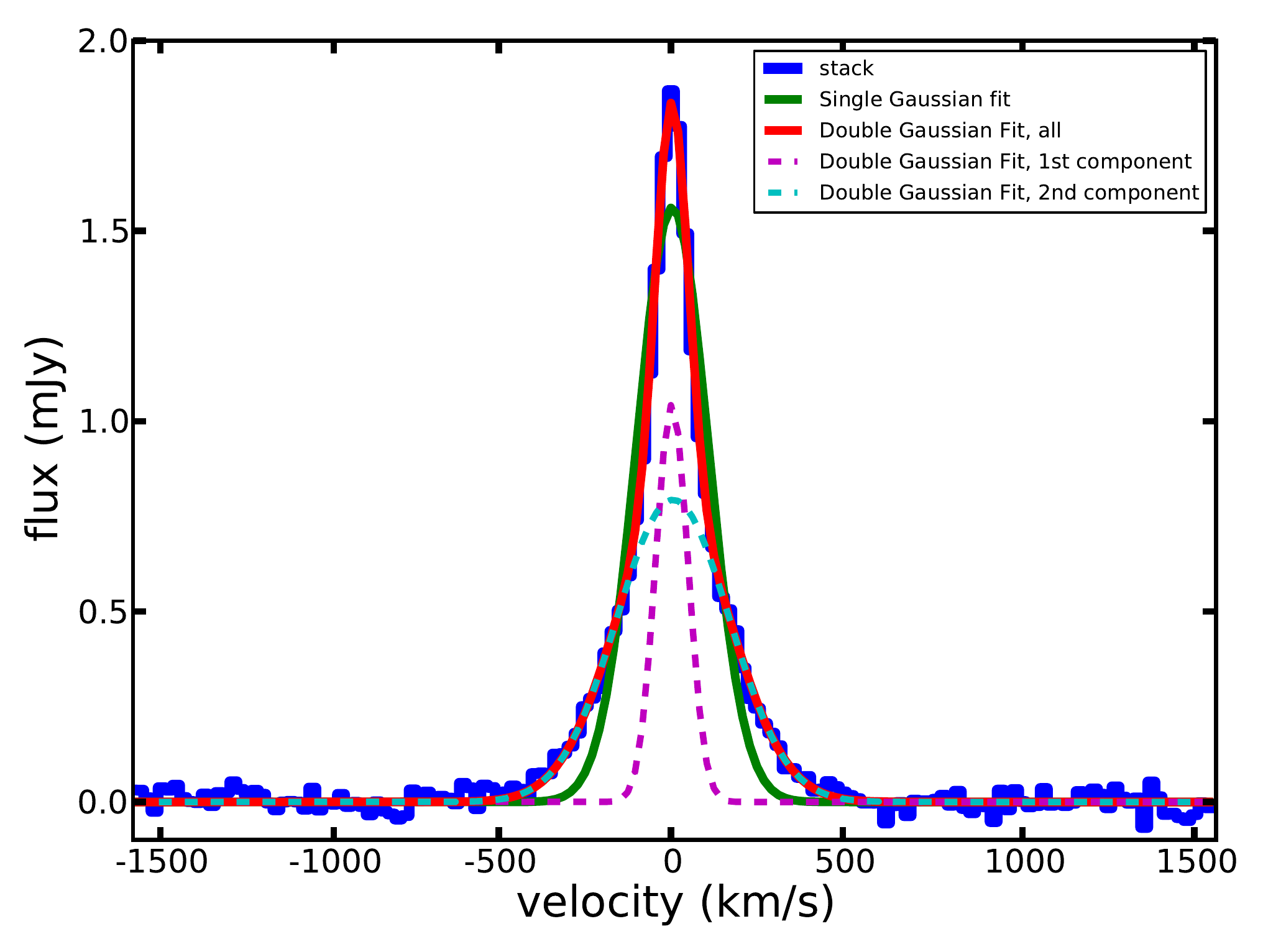}
\centering
\caption{Original stack, before rebinning. Fitted with one (green line) and two (red line) Gaussians.}
\label{fig:noRebinFit}
\end{figure}

All cubes are stacked using mean stacking. The resulting spectrum can be seen on Figure \ref{fig:noRebinFit}. Here the spectrum is extracted from the stacked cube by summing all $8\times8$ pixels of the stacked cube. One can see that the resulting stack is better fitted by two Gaussian line profiles, even if the lines where originally simulated with single Gaussian profiles.

\begin{figure}
\includegraphics[width=\linewidth]{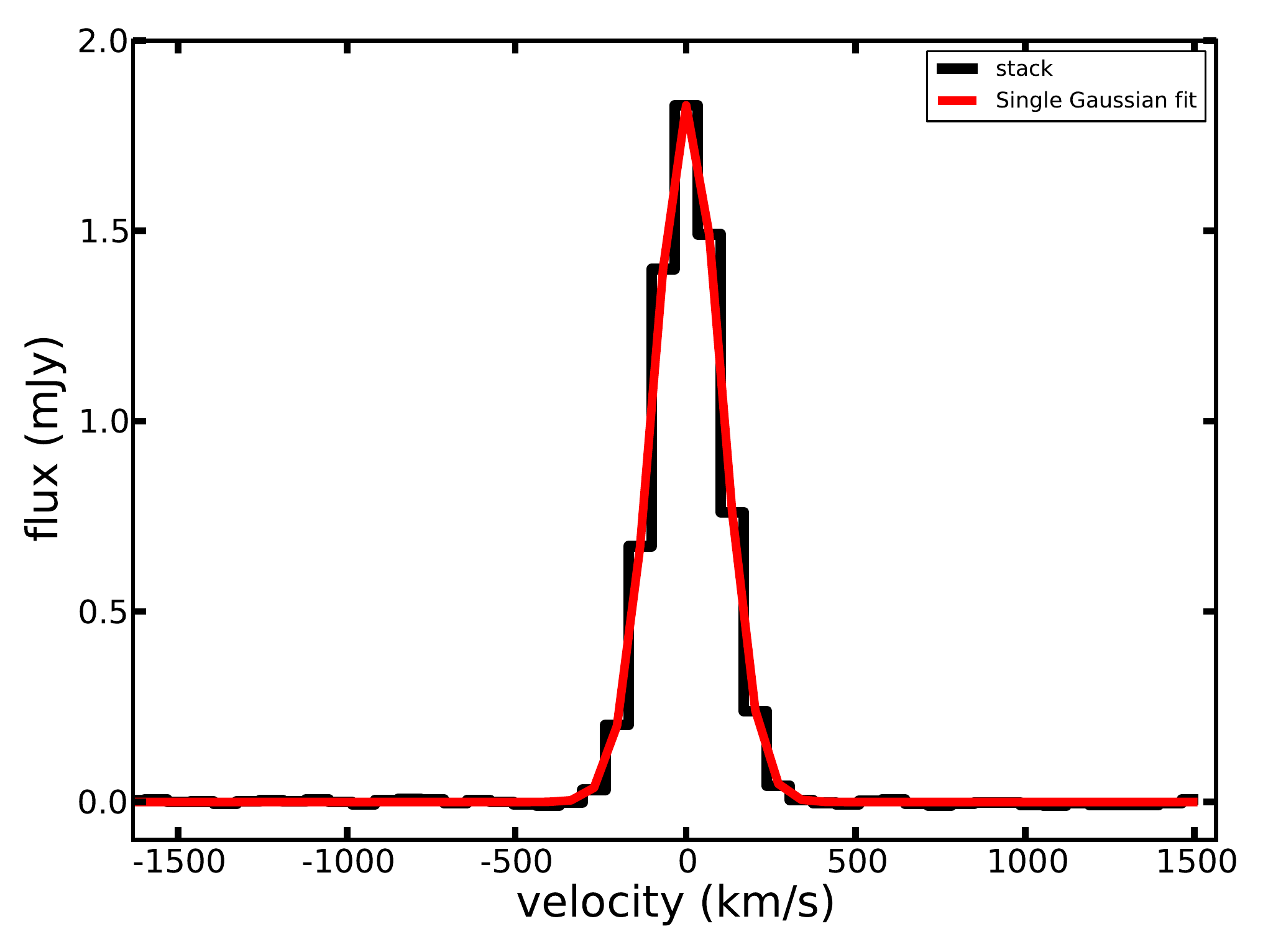}
\centering
\caption{Stack result after rebinning. Shows very good agreement with single component Gaussian fitting.}
\label{fig:RebinFit}
\end{figure}

To avoid this effect {\it spectral rebinning} is performed on the cubes, using the method embedded in \textsc{LineStacker}. The method allows for automated fitting to estimate the line widths, which are used for the rebinning. The rebinned cubes are stacked again, using mean stacking. The stacking result is presented on Figure \ref{fig:RebinFit}, and one can see that, after rebinning, the resulting stack spectrum is purely Gaussian, as expected. {The width of the stacked line is $\sim$\,242km\,s$^{-1}$ (to be compared to the average line width before stacking: $\sim$\,231km\,s$^{-1}$).}

\subsection{1D extended example} \label{ex:1Dextended}

Because averaged properties of the studied population are retrieved from stacking, the underlying assumption in stacking data is that all sources have similar properties. In some cases, it is known that not all sources will show similar line properties, but it is not known a priory which sources. This can be well illustrated when searching for outflows, where orientation could affect the projected kinematics properties and thereby the line width \citep[e.g.][]{Stanley2019}. 

\begin{figure}
\includegraphics[width=\linewidth]{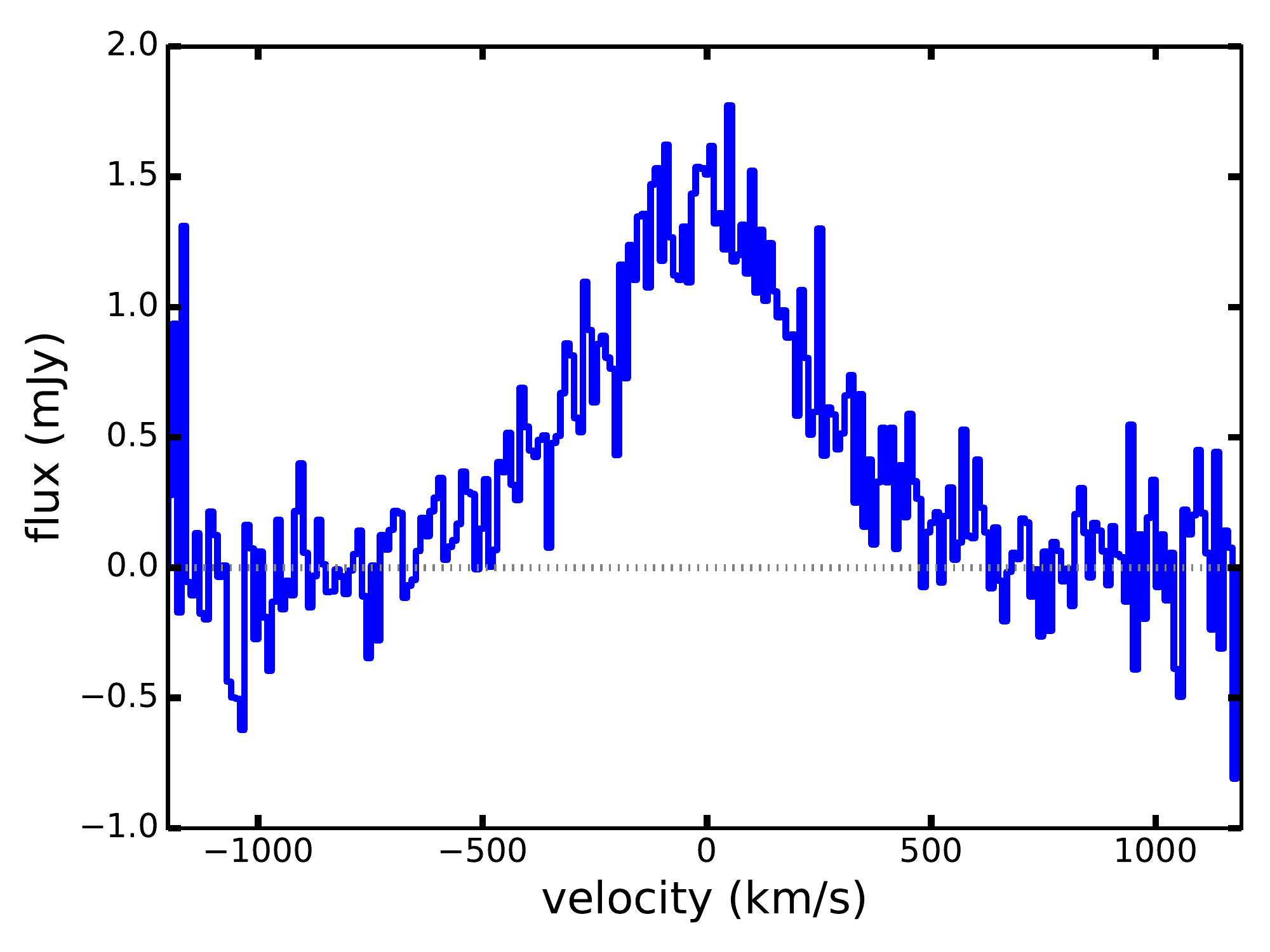}
\centering
\caption{Median stack of the 50 spectra with non homogeneous line amplitude.}
\label{fig:1DextendedStack}
\end{figure}

In this example we present the use of bootstrapping and subsampling. Here we will be using the 1D module of \textsc{LineStacker}, and hence stack spectra directly, and not whole cubes. The data-set consists of 50 spectra. The spectra are composed of a Gaussian line, randomly centered, and some noise. The amplitude of the Gaussian line is typically of order of the noise, however, 10 of the sources have an average line amplitude 10 times higher. Such an inhomogeneous distribution has been chosen to showcase the performance and usage of both bootstrapping and subsampling. All spectra are stacked using median stacking, the stacking result is shown on Figure \ref{fig:1DextendedStack}. 

\begin{figure}
\includegraphics[width=\linewidth]{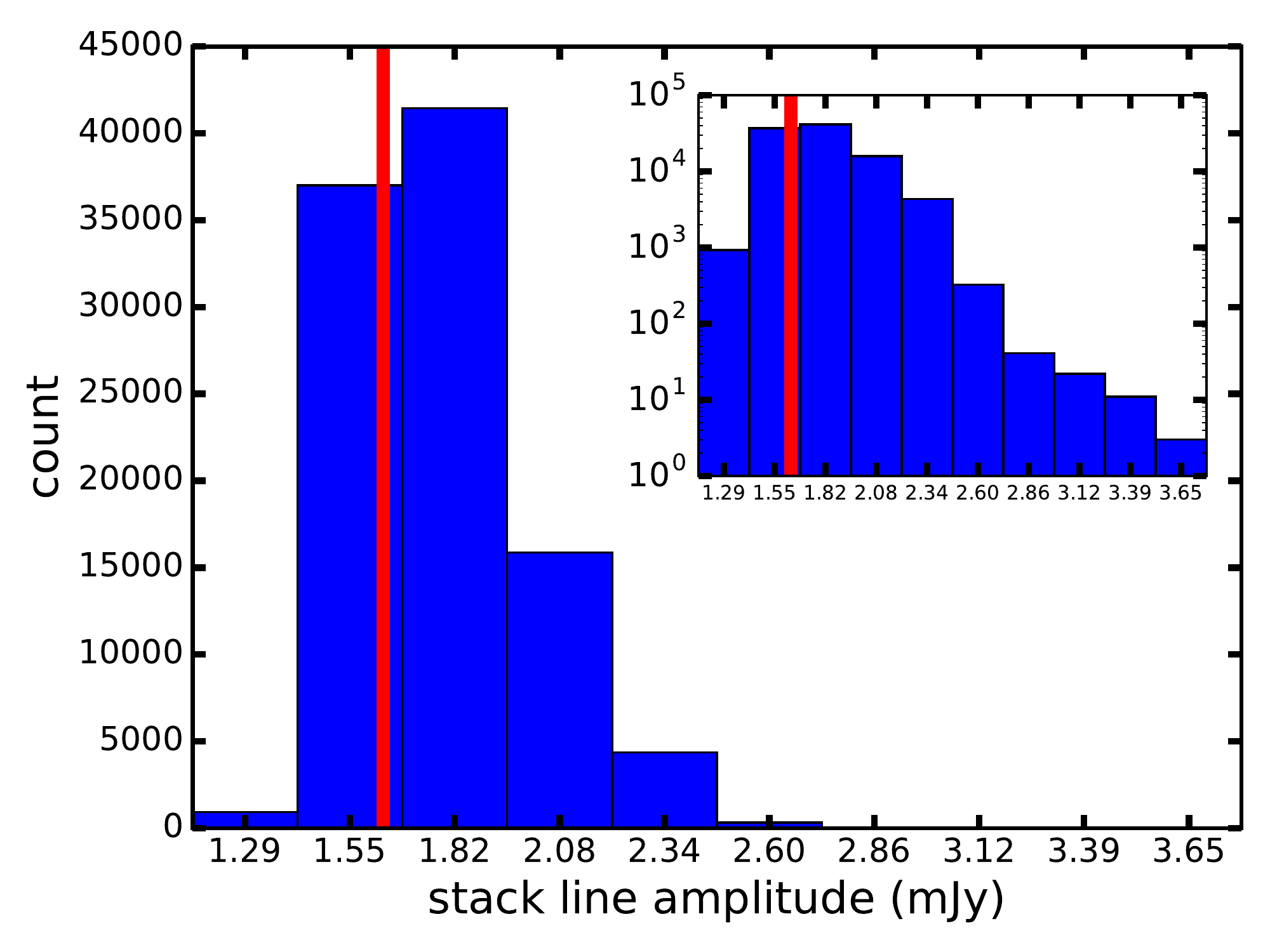}
\centering
\caption{Distribution of the results from the median bootstrap analysis. The straight red line indicates the result from the stack of the entire sample.}
\label{fig:boot}
\end{figure}

The second step is to show the usage of bootstrapping methods. Here we perform bootstrapping using median stacking, and iterating 100000 times. Results can be seen on Figure \ref{fig:boot}.

The bootstrapping results show a clear skewed distribution toward results of higher amplitude. This typically implies an inhomogeneous distribution of the sources amplitude. While a similar conclusion could have been drawn from looking at bootstrapping paired with mean stacking, it would not have been as easy to display, which is why we chose to show median stacking in this example. When using bootstrapping paired with mean stacking, the in-homogeneity of the results can be typically deduced from the width of the bootstrap distribution -- showing a much larger spread of the results than it would have if the sample was homogeneous. 

\begin{figure*}
    \includegraphics[width=\textwidth]{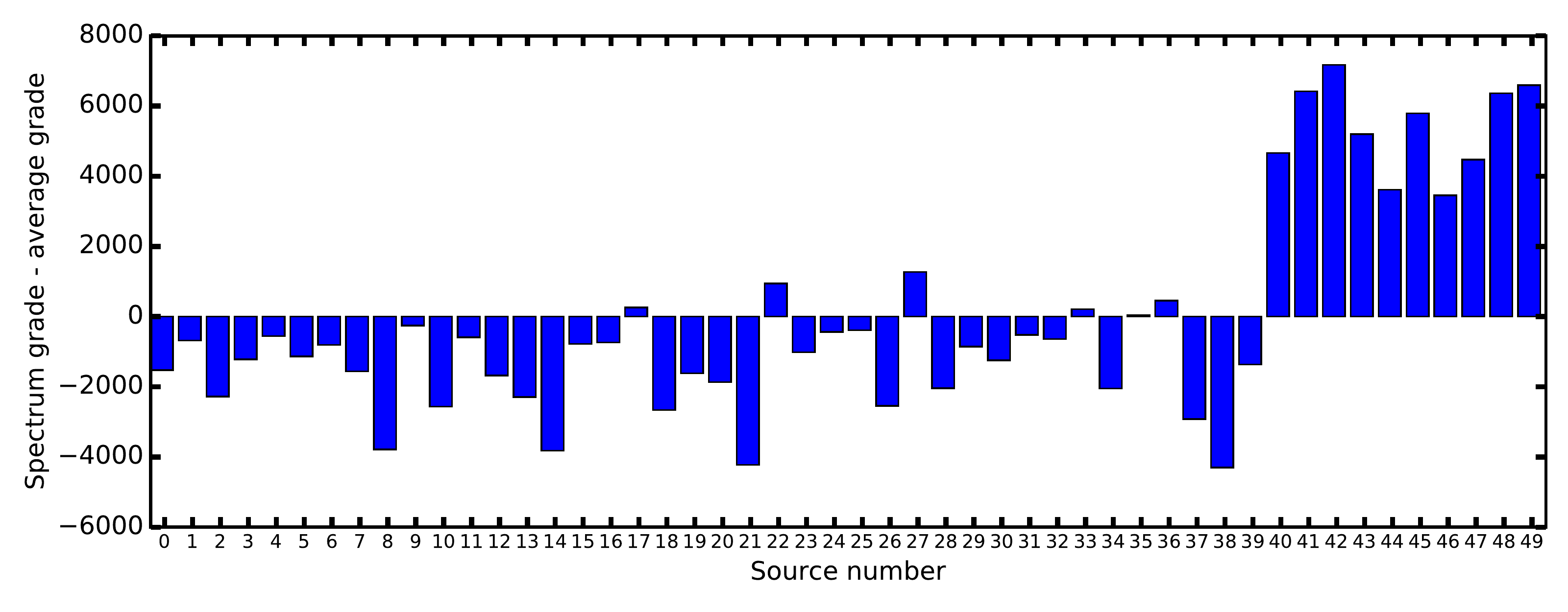}
    \caption{Results from subsampling analysis: the average grade is subtracted from each sources grade.}
\label{fig:subSample1D}
\end{figure*}

Since bootstrapping allowed to suspect the presence of outliers in the data, one can now use subsampling to try to identify them. Because the bootstrapping analysis indicated a skewed distribution of the lines amplitude, the amplitude will be used as a criteria to separate the subsamples (see Section \ref{subsample} for a description of the usage of grades in our subsampling method). The subsampling method is performed 100000 times and its results can be seen of Figure \ref{fig:subSample1D}.

The distribution of source grade shows a clear in-homogeneity, and it is easy here to identify the last 10 sources as having higher amplitude.

\label{lastpage}
\end{document}